\pdfoutput=1
\documentclass[12pt]{article}
\usepackage{amssymb,amsmath}
\usepackage{color,graphicx}
\usepackage{cite}
\usepackage{multirow}
\usepackage{scrextend}
\usepackage{hyperref}
\addtokomafont{labelinglabel}{\sffamily\bfseries}
\newcommand{\beq} {\begin{equation}}
\newcommand{\eeq} {\end{equation}}
\newcommand{\bear}{\begin{eqnarray}}
\newcommand{\eear}{\end{eqnarray}}
\newcommand{\be}{\begin{equation}}
\newcommand{\ee}{\end{equation}}
\newcommand{\bea}{\begin{eqnarray}}
\newcommand{\eea}{\end{eqnarray}}
\newcommand{\im}{{\rm Im}}
\newcommand{\re}{{\rm Re}}

\numberwithin{equation}{section}

\begin{document}
 
\begin{flushright}
HIP-2018-30/TH
\end{flushright}

\begin{center}

\centerline{\Large {\bf Information flows in strongly coupled ABJM theory}}

\vspace{8mm}

\renewcommand\thefootnote{\mbox{$\fnsymbol{footnote}$}}
Vijay Balasubramanian,${}^{1,2}$\footnote{vijay@physics.upenn.edu}
Niko Jokela,${}^{3,4}$\footnote{niko.jokela@helsinki.fi} \\
Arttu P\"onni,${}^{3,4}$\footnote{arttu.ponni@helsinki.fi} and
Alfonso V. Ramallo${}^{5,6}$\footnote{alfonso@fpaxp1.usc.es}

\vspace{6mm}

${}^1${\small \sl David Rittenhouse Laboratory, University of Pennsylvania} \\
{\small \sl PA 19104 Philadelphia, USA} \\

\vspace{2mm}
${}^2${\small \sl Theoretische Natuurkunde, Vrije Universiteit Brussel} and {\small \sl International Solvay Institutes} \\
{\small \sl Pleinlaan 2, B-1050 Brussels, Belgium} \\

\vspace{2mm}
${}^3${\small \sl Department of Physics} and ${}^4${\small \sl Helsinki Institute of Physics} \\
{\small \sl P.O.Box 64, FIN-00014 University of Helsinki, Finland} 

\vspace{2mm}
${}^5${\small \sl Departamento de F\'isica de Part\'iculas} and ${}^6${\small \sl Instituto Galego de F\'isica de Altas Enerx\'ias} \\
{\small \sl Universidade de Santiago de Compostela, E-15782 Santiago de Compostela, Spain}

\end{center}

\vspace{6mm}

\setcounter{footnote}{0}
\renewcommand\thefootnote{\mbox{\arabic{footnote}}}

\begin{abstract}
\noindent
We use holographic methods to characterize the RG flow of quantum information in a Chern-Simons theory coupled to massive fermions.    First, we use entanglement entropy and mutual information between strips to derive the dimension of the RG-driving operator and a monotonic c-function. We then display a scaling regime where, unlike in a CFT, the mutual information between strips changes non-monotonically with strip width, vanishing in both IR and UV but rising to a maximum at intermediate scales.  The associated information transitions also contribute to non-monotonicity in the conditional mutual information which characterizes the independence of  neighboring strips after conditioning on a third.   Finally, we construct a measure of extensivity which tests to what extent information that region A shares with regions B and C is additive.  In general, mutual information is super-extensive in holographic theories, and we might expect super-extensivity to be maximized in CFTs since they are scale-free.  Surprisingly, our massive theory is more super-extensive than a CFT in a range of scales near the UV limit, although it is less super-extensive than a CFT at all lower scales.  Our analysis requires the full ten-dimensional dual gravity background, and the extremal surfaces computing entanglement entropy explore all of these dimensions.
\end{abstract}

\newpage
\tableofcontents
\newpage


\section{Introduction}

Interactions in a quantum field theory (QFT) cause quantum information to be non-locally distributed across space through the phenomenon of entanglement in the wavefunction.    It is of great interest to understand how this structure of shared information changes with energy scale, or equivalently, during renormalization group (RG) flow.   Essentially the only  tool we have to study this question in strongly coupled field theories is holography.  Even in these settings there is a key challenge: there are very few exact solutions in holographic gravity that correspond to non-trivial renormalization group flows in QFT.   

Many ``phenomenological'' holographic treatments of RG flow are constructed so that a $d$-dimensional field theory is described by a $(d+1)$-dimensional theory of gravity in which the extra dimension represents QFT scale.  Gravitational solutions can be easier to obtain in this simplified setting. However, in cases where an actual gauge-gravity duality has been constructed in string theory, there are always the ten dimensions of string theory in the gravitational description, with internal symmetries of the QFT represented by the structure of the extra dimensions beyond the radial direction which represents QFT scale.  In such settings which are fully grounded in string theory, RG flows of QFT are described as complete ten-dimensional gravitational backgrounds where the ``internal'' and ``AdS'' dimensions interact, mix and sometimes exchange roles.  Finding exact gravitational solutions in this complete setting is difficult.    Our goal in this paper is to examine the RG flow of quantum information in a strongly coupled QFT running between non-trivial interacting fixed points, in a scenario where we have the powerful lever of a complete ten-dimensional classical gravitational dual.

To this end, we focus on the ABJM theory \cite{Aharony:2008ug} which is the $(2+1)$-dimensional analogue of the 4d maximally supersymmetric Yang-Mills theory. This 3d Chern-Simons gauge theory has a symmetry group $U(N)_k\times U(N)_{-k}$  with levels $(k,-k)$ coupled to matter fields in the bifundamental representation of the gauge group.  When $N$ and $k$ are  large, the model has a dual holographic description as an  $AdS_4\times {\mathbb{CP}}^3$ geometry with fluxes in Type IIA supergravity.  The ABJM theory can be generalized by adding fundamental quarks transforming in the $(N,1)$ and $(1,N)$ representations of the gauge group \cite{Hohenegger:2009as,Gaiotto:2009tk}.  In the gravity dual, these quarks are incorporated by adding D6-branes wrapping an ${\mathbb{RP}}^3$ inside the internal ${\mathbb{CP}}^3$ and extended along $AdS_4$.

When the number of flavors $N_f$ is sufficiently small, the field theory can be treated in a quenched approximation where the loops of fundamentals are suppressed.    In the same limit,  the D6-branes in the gravity dual can be treated as probes in the  $AdS_4\times {\mathbb{CP}}^3$  background.  When $N_f$ is larger so that the field theory is not quenched, the D6-branes in the gravity description will backreact on the geometry.   Constructing the backreacted supergravity solution is a difficult task in general.  However, in the Veneziano limit \cite{Veneziano:1976wm} in which both the number of colors $N$ and flavors $N_f$  are taken to be large with  $N_f/N$ fixed,  we can employ a systematic perturbative approximation \cite{Nunez:2010sf}. Using this  technique one can find the complete gravity duals.  In our context, these generalized ABJM geometries were found in \cite{Conde:2011sw,Jokela:2012dw,Bea:2014yda} for  massless flavors and in \cite{Bea:2013jxa} for  massive flavors, with a recent extension \cite{Bea:2017iqt} to a non-commutative geometry. 
This is not what some call a massive deformation of the ABJM theory \cite{Hosomichi:2008jb,Gomis:2008vc}; entanglement flows in that context are studied in \cite{Kim:2014yca,Kim:2014qpa}.   The extremal surfaces that holographically compute entanglement in our setting will explore all 10 dimensions of the dual gravity theory (see previous examples in, {\emph{e.g.}}, \cite{Jones:2016iwx,Balasubramanian:2017hgy}).

\section{Background solution} \label{sec:background}

The maximally supersymmetric three-dimensional $U(N)_k\times U(N)_{-k}$ Chern-Simons theory coupled to massive bifundamental quarks should be conformal in the UV, at scales well above the quark masses, and in the IR, at scales well below the quark masses.  Thus, we expect a dual description in Type IIA supergravity that tends to $AdS_4$ times a compact factor both near infinity and in the deep interior.   The relevant solution, interpolating smoothly between  two $AdS_4$ spacetimes,  was found in \cite{Bea:2013jxa}.  The string frame metric in this solution is:
\be\label{eq:metric}
 ds_{10}^2 = h(x)^{-1/2}dx^2_{1,2}+h(x)^{1/2}e^{2g(x)}\left[\frac{dx^2}{x^2}+q(x)ds^2_{\mathbb{S}^4}+ds^2_{\mathbb{S}^2_{f} }\right]  \ ,
\ee
where $ds^2_{1,2}$ in the Minkowski metric in $2+1$ dimensions,  $ds^2_{\mathbb{S}^4}$ is the standard metric of a unit four-sphere,  
and $\mathbb{S}^2_{f}$ is a fibered two-sphere (see below).   We are just presenting the metric here and not the fluxes, since the latter do not play a role in our analyses.
The radial coordinate $x$ in (\ref{eq:metric}) is related to a canonical coordinate $r$ in 
which the metric is asymptotically $ds^2 = -r^2 dt^2 + dr^2/r^2 + \ldots$
as follows
\be\label{eq:rtox}
 dr= e^{g}\frac{dx}{x}  \ .
\ee
Through this relation, the IR $(r=0)$ and the UV $(r=\infty)$ correspond to $x=0$ and $x=\infty$, respectively.   As the geometry is not globally $AdS$, the dilaton is also not a constant (see below), but in the IR and UV the geometry tends towards AdS$_4$ in Poincar\'{e} coordinates times a compact factor.  Thus $x=0$ is the Poincar\'{e} horizon.

The function $e^{2g(x)}$, the local radius of the internal $\mathbb{S}^2$, quantifies the relative squashing between the internal and external manifolds. The radius of $\mathbb{S}^4$ is given by the function $e^{2f}:=q \,e^{2g}$. The function $q(x)$ measures the relative squashing between the $\mathbb{S}^2$ fiber and the $\mathbb{S}^4$ base on the internal manifold. A convenient representation of the metric for $\mathbb{S}^4$ is
\be\label{eq:S4metric}
 ds^2_{\mathbb{S}^4}  = \frac{4}{\left(1+\xi^2\right)^2}\left[d\xi^2+\xi^2\sum_{i=1}^3(\omega^i)^2 \right] \ ,
\ee
where $\omega^i$ are the $SU(2)$ left-invariant one-forms which satisfy $d\omega^i=\frac{1}{2}\epsilon_{ijk}\omega^j\wedge\omega^k$ and $0\leq\xi<\infty$.
The metric of the fibered two-sphere $\mathbb{S}^2$ can be written in terms of one-forms $E^1,E^2$:
\bea
 ds^2_{\mathbb{S}^2_{f}} & = & \left(E^1\right)^2+\left(E^2\right)^2 \\
 E^1 & = & d\theta+\frac{\xi^2}{1+\xi^2}\left(\sin\varphi\omega^1-\cos\varphi\omega^2\right) \\
 E^2 & = & \sin\theta\left(d\varphi-\frac{\xi^2}{1+\xi^2}\omega^3\right)+\frac{\xi^2}{1+\xi^2}\cos\theta\left(\cos\varphi\omega^1+\sin\varphi\omega^2\right)    \ ,
\eea
where $0\leq \theta\leq\pi,0\leq\varphi < 2\pi$.

The massive quarks in the field theory require the inclusion of D6-branes that span the AdS$_4$ part of the geometry and wrap a part of the internal manifold.     The D6-brane charge density is measured by violation of the Bianchi identity of the RR two-form $F_2$, and this turns out to be proportional to $\eta-1$ \cite{Bea:2013jxa} where
\be\label{eq:eta}
 \eta(x) = 1+\hat\epsilon \left(1-\frac{1}{x^2}\right)\Theta(x-1) \ ,
\ee
in terms of the Heaviside function $\Theta(z)$.
Thus, $\eta$ dictates the distribution of smeared flavor D6-branes that act as sources for the SUGRA equations of motion. Notice that $\eta=1$ for $x<1$, meaning that the D6-branes do not span the whole radial coordinate but end above the Poincar\'e horizon. Thus, in the following, we call the region $x\leq 1$  the ``cavity''. The parameter $\hat\epsilon$ encodes backreaction of the massive quarks and relates to the number $N_f$ of D6-branes as:
\be
 \hat\epsilon = \frac{3N_f}{4k} = \frac{3}{4}\frac{N_f}{N}\lambda \ ,
 \label{epsilonhatdef}
\ee
where the 't Hooft coupling is related to the Chern-Simons level $k$ and the number of colors $N$ as $\lambda = N/k$.

The metric ansatz (\ref{eq:metric}) is guaranteed to preserve $\mathcal{N}=1$ supersymmetry in three dimensions \cite{Conde:2011sw,Bea:2013jxa}, if the following master equation is satisfied
\be\label{eq:masterequation}
 W'' +4\eta'+\left(W'+4\eta\right)\left[\frac{W'+10\eta}{3W}-\frac{W'+4\eta+6}{x\left(W'+4\eta\right)}  \right]=0
\ee
for $W$ defined as 
\be\label{eq:Wdefinition}
 W(x) = \frac{4}{k}h^{1/4}e^{2f-g-\phi} \ .
\ee
Then, up to freedom in integration constants, the functions $f$, $g$, $h$, and the dilaton $\phi$ can be constructed from $W$ (see Appendix~\ref{app:details}). Moreover,  the squashing function $q$ can be written in terms of the master function $W$ and its derivative as:
\beq
q\,=\,{3 W\over x(W'+4\eta)}\,\,.
\label{q_W_Wprime}
\eeq

The quark mass $m_q$ encoded in this gravitational background can be computed by evaluating the Nambu-Goto action of a fundamental string stretched along the holographic direction between $x=0$ and $x=1$ at fixed values of the other spacelike coordinates.  This gives $m_q=r_q/2\pi\alpha'$, where $\alpha'$ is the Regge slope (which we will set to 1), with $r_q$ given by:
\be\label{eq:rq}
 r_q = \int_0^1 dx \frac{e^g}{x} \ .
\ee
Following (\ref{eq:rtox}), $r_q$ is just the canonical $r$ coordinate of the tip of the flavor D6-branes.

\paragraph{The IR limit: }   In the region $x<1$ (the ``cavity'' region) there are no D6-branes (see (\ref{eq:eta})).  Thus, the solution in this region should solve the matter-free equations with $N_f \to 0 $,  $\hat\epsilon \to 0$, and $\eta \to 1$.   In this limit, $W=2x$ gives one solution to the master equation. This choice reproduces the  original matter-free ABJM solution after transforming to the canonical coordinate $r$ \cite{Bea:2013jxa}. In this case the squashing function $q$  in (\ref{q_W_Wprime}) becomes equal to one and the internal metric in (\ref{eq:metric}) is the standard Fubini-Study metric of ${\mathbb C}{\mathbb P}^3$ giving a 10d metric
\beq
ds^2_{10, \epsilon=0}\,=\,L_{ABJM}^2\,ds^2_{AdS_4}\,+\,4\,L_{ABJM}^2\,ds^2_{{\mathbb C}{\mathbb P}^3}\,\,,
\label{ABJM-metric}
\eeq
where $ds^2_{AdS_4}$ and $ds^2_{{\mathbb C}{\mathbb P}^3}$ are respectively
 the  $AdS_4$  and ${\mathbb C}{\mathbb P}^3$ metrics. The former,  in Poincar\'e coordinates,  is given by:
\beq
ds^2_{AdS_4}\,=\,r^2\,d x_{1,2}^2\,+\,{dr^2\over r^2} \ .
\label{AdS4metric}
\eeq
 In (\ref{ABJM-metric})  $L_{ABJM}$ is the radius of the $AdS_4$ part of the metric and  in string units is 
\be
L_{ABJM}^4\,=\,2\pi^2\,{N\over k}\,\,,
\label{ABJM-AdSradius}
\eeq
where $N$ and $k$  are two integers which correspond, in the gauge theory dual, to the rank of the gauge groups and the Chern-Simons level, respectively.  

For our purpose of solving the full equations with matter, the pure ABJM solution does not suffice in the $x<1$ region.  Rather we need a solution that will match appropriately onto the solution in the $x>1$ region where the D6-branes backreact on the geometry.   Fortunately, in the limit $\hat\epsilon\to 0$, there is a one-parameter family of analytic solutions to (\ref{eq:masterequation}):
\be\label{eq:running}
 W_{running} = \frac{4(1+4\gamma x)x}{1+\sqrt{1+4\gamma x}} \ ,
\ee
where $\gamma$ is a constant. These ``running solutions''  reduce to ABJM in the deep IR ($W_{running}(x\to 0)\to 2 x$).    But as $x$ increases, the running solution flows away from the ABJM fixed point and eventually asymptotes at $x\to\infty$ to a metric describing a resolved Ricci flat cone.     With a finite number of quark flavors and hence $\hat\epsilon > 0$, the running solution will apply in the ``cavity'' $x < 1$ which is not penetrated by the D6-branes and hence is locally described by an $N_f = 0$ solution.  We will match the running solution in the $x<1$ region onto the appropriate metric in the $x>1$ region.

\paragraph{The UV limit: }  To gain control over the $x > 1$ region, we first consider the asymptotic UV limit $x\to\infty$.  In this limit,  the D6-brane distribution tends to a constant since $\eta\to 1+\hat\epsilon\equiv \eta_0$.  Then one solution of the master equation (\ref{eq:masterequation}) is
\be\label{eq:flavoredABJM}
 W(x\to\infty) \approx \frac{q_0(\eta_0+q_0)}{2-q_0}x \ ,
\ee
where the squashing function $q$ is also a constant $q_0 = \left( 3+3\eta_0-\sqrt{9\eta_0^2-2\eta_0+9} \right)/2$. 
The resulting ten-dimensional asymptotic geometry is of the type $AdS_4\times {\cal M}_6$, where 
${\cal M}_6$ is a squashed version of  ${\mathbb C}{\mathbb P}^3$ with constant squashing factors. Indeed, the metric takes the form
\beq
ds_{10}^2\,\approx\,L_0^2\,\,ds^2_{AdS_4}\,+\,
{L_0^2\over b^2}\,\Big[\,q_0\,ds^2_{{\mathbb S}^4}\,+\,
ds^2_{\mathbb{S}^2_{f} }\,\Big]\,\,,
\label{metric-AdS-asymp}
\eeq
where $L_0$ is constant and given by
\beq
L_0^4\,=\,128\pi^2\,{N\over k}\,
{(2-q_0)\,q_0^3\over (\eta_0+q_0)\,(q_0+1)^5}\,\,,
\label{L0_explicit}
\eeq
and $b$ is another constant, which can be written in terms of the asymptotic UV squashing $q_0$ as
\beq
b\,=\,{2\,q_0\over q_0+1}\,\,.
\label{b_new}
\eeq
The parameter $b$ will play an important role in what  follows. Its interpretation is clear from  (\ref{metric-AdS-asymp}): it represents the relative squashing of the  ${\mathbb C}{\mathbb P}^3$ part of the asymptotic metric with respect to the $AdS_4$  part. In the unflavored case $\hat \epsilon=0$ we have $q_0=b=1$. In general $q_0$ and $b$ grow with $\hat\epsilon$ and reach their maximal  values $q_0=5/3$ and $b=5/4$ when 
$\hat\epsilon\to\infty$.

In fact, (\ref{eq:flavoredABJM}) solves the master equation for all $x$ if $\eta$ is constant everywhere, and corresponds to the limit of {\emph{massless}} quarks ($m_q \propto r_q\to 0$) that backreact in the AdS description.  This case is discussed in depth in \cite{Conde:2011sw,Jokela:2012dw,Jokela:2013qya,Bea:2014yda}, and gives the desired asymptotics because at energies far above the mass scale of the quarks they should act as if they are massless.  Holographically, this means that the full solution should asymptote to (\ref{eq:flavoredABJM}).

\paragraph{The full solution: }
The solution to the master equation that we are interested in has two pieces. Inside the cavity $(x\leq 1)$ the master function $W$ is given by (\ref{eq:running}). At the boundary of the cavity $x=1$ we need to glue this solution continuously to another one which asymptotes to (\ref{eq:flavoredABJM}). This can be obtained numerically by the standard shooting technique starting at $x=1$, with boundary conditions $W(1)$ and $W'(1)$ given by the running solution (\ref{eq:running}). Shooting is done by varying the constant $\gamma$ in the running solution for $x<1$ until the desired asymptotic solution for $W(x\to \infty)$ is obtained by integrating outward with boundary conditions that impose continuous $W$ and $W'$ at $x=1$. A generic value of $\gamma$ will provide a solution for which $W\sim x^{3/2}$ as $x\to \infty$, {\emph{i.e.}}, not asymptotically $AdS_4$ but rather a $G_2$ cone. For a given $\hat\epsilon$ there exists a unique $\gamma=\gamma(\hat\epsilon)$ such that $W\sim x$, as $x\to\infty$, {\emph{i.e.}}, such that the solution asymptotes to (\ref{eq:flavoredABJM}). Such a piecewise metric then describes an interpolation between two $AdS_4$ spacetimes with different radii.   These solutions were constructed numerically in  \cite{Bea:2013jxa} and  as a power series in $\hat\epsilon$ in \cite{Bea:thesis}.  In addition to the numerical solutions, we will also use this power series, which is reviewed in Appendix~\ref{app:details}.

\subsection{Holographic Callan-Symanzik equation}

The solution given above is a holographic description of the renormalization group flow of a strongly interacting theory of matter and gauge fields.   The AdS$_4$ geometries at $x \to 0$ and $x \to \infty$ indicate that the flow runs between different conformal fixed points.  We can characterize the flow in terms of a Callan-Symanzik equation for the quark mass, extending the discussion in \cite{Jokela:2013qya}. 

The natural mass scale in a top-down holographic setup such as ours involving flavor and color branes is the distance between these two sets of branes. This distance is related to the mass of fundamental quarks in the field theory that are dual to the open strings ending on the flavor brane \cite{Karch:2002sh}. In our model, the flavor branes are D6-branes embedded supersymmetrically in the background. The quark mass is obtained by evaluating  the Nambu-Goto action  for a string extended from the origin and ending on the branes \cite{Erdmenger:2007cm}. 

To evaluate the quark mass, consider a probe D6-brane embedded along the four $AdS_4$ directions and wrapping a three-dimensional cycle inside the compact internal manifold. The precise form of this cycle for a supersymmetric embedding can be determined by using kappa symmetry. It was shown in \cite{Conde:2011sw} that this  requires that the pullbacks to the worldvolume of the $SU(2)$ one-forms $\omega^1$ and $\omega^2$ of (\ref{eq:S4metric}) must vanish. Extending the brane along the coordinate $\varphi$ of the ${\mathbb S}^2$ fiber, the induced metric on the brane worldvolume takes the form:
\bea
   ds^2_7 & = & \frac{1}{\sqrt{h(r)}} dx_{1,2}^2 + \sqrt{h(r)}(1+e^{2g(r)}\theta'(r)^2)dr^2 \nonumber \\
   & & + e^{2g(r)} \sqrt{h(r)} \left( q(r) d\alpha^2 + q\sin^2\alpha d\beta^2 + \sin^2\theta \left( d\psi + \cos\alpha d\beta \right)^2 \right)\,\,,
\eea
where, for convenience, we are using the canonical $AdS_4$ coordinate $r$ introduced in (\ref{eq:rtox}). The angular coordinates $\alpha$, $\beta$, and $\psi$ are related to $\xi$, $\varphi$, and to the one used to parameterize the pullback of $\omega^3$ (see \cite{Conde:2011sw} for details). The kappa symmetry condition then reduces to the following first-order differential equation for $\theta=\theta(r)$:
\beq
e^{g(r)}\,{d\theta\over dr}\,=\,\cot\theta\,\,.
\eeq
This equation can be integrated exactly after a change of variable to a new radial coordinate $u$:
\begin{align}
   \frac{dr}{e^{g(r)}} = \frac{du}{u} \quad \rightarrow \quad u = \exp\left( \int^r e^{-g(r')} dr \right) \ .
   \label{r_u}
\end{align}
Then, the embedding function $\theta$ is simply
\beq
\cos\theta\,=\,{R_0\over u} \ ,
\eeq
where $R_0$ is an integration constant. Thus, for our SUSY embedding $u\cos\theta$ is constant. In order to easily describe the corresponding seven-dimensional surface, it is convenient to introduce two new coordinates $(R, \rho)$ as
\begin{equation}
   R = u \cos\theta ~~~~;~~~~
   \rho = u \sin\theta \ .
\end{equation}
In these coordinates the $(r,\theta)$ sector of the ten-dimensional metric takes the form
\begin{align}
   \sqrt{h(r)} \left( dr^2 + e^{2g(u)} d\theta^2 \right) = \frac{\sqrt{h(u)} e^{2g(u)}}{u^2} \left( dR^2 + d\rho^2 \right)\ ,
\end{align}
where $u^2=R^2+\rho^2$. For supersymmetric embeddings $R=R_0={\rm constant}$, which leads to the induced worldvolume metric
\bea
   ds^2_7 & = & \frac{1}{\sqrt{h}} dx_{1,2}^2 + {\sqrt{h}\,e^{2g}\over \sqrt{R_0^2+\rho^2}}\,d\rho^2 \nonumber \\
   & & + e^{2g} \sqrt{h} \left( q d\alpha^2 + q\sin^2\alpha d\beta^2 + \sin^2\theta \left( d\psi^2 + \cos\alpha d\beta \right)^2 \right) \ ,
\eea
where all functions depend on $\rho$ through the combination $\sqrt{R_0^2+\rho^2}$. Therefore, $\rho$ plays the role of the holographic coordinate on the worldvolume. Actually, in the UV region $\rho\approx u\to\infty$ we have that $e^{g}\approx r/b$ and  (\ref{r_u}) can be integrated as $r\sim \rho^{{1\over b}}$. Since $r$ is the canonical $AdS_4$ coordinate we should identify the energy scale with $\rho^{{1\over b}}$.

Consider now a fundamental string located at the point $\rho=\rho_*$. The string extends from $R=0$ to its boundary value $R=R_0$, where it intersects with the D6-brane. The running mass of the corresponding dual (valence) quark is equal to the Nambu-Goto action of this string per unit time.  The induced metric on the string worldsheet extended in $(t,R)$ at $\rho=\rho_*$ is given by
\begin{align}
   ds_2^2 = - \frac{dt^2}{\sqrt{h}} + \sqrt{h}\, e^{2g} \frac{dR^2}{R^2+\rho_*^2} \ .
\end{align}
The running quark mass is then
\be
   m_q = \frac{1}{2\pi\,\alpha'^{3/2}} \int_0^{R_0} \sqrt{-\det g_2} \,dR = \frac{1}{2\pi\alpha'^{3/2}} \int_0^{R_0} \frac{e^{g(u)}dR}{\sqrt{R^2+\rho_*^2}} \ .
  \label{running_mass}
\ee

We want to find an evolution equation for $m_q$ with respect to the energy scale $\Lambda$. As argued above, we should identify the energy scale with $\rho_*^{1/b}$. Therefore, we define:
\begin{align}
   \Lambda := \rho_*^{1/b}\ .
\end{align}
We can work out the evolution equation for $m_q$, which after a straightforward calculation boils down to:
\be
    2\pi\alpha'^{3/2} \frac{\partial m_q}{\partial \log\Lambda} = b\rho_* \frac{\partial m_q}{\partial \rho_*} = b \int_0^{R_0} {\sqrt{R^2+\rho_*^2}\over R}\,\partial_{R}\,e^{g}\,dR - \frac{b R_0 e^{g(\sqrt{R_0^2+\Lambda^{2b}})}}{\sqrt{R_0^2+\Lambda^{2b}}}\ .
    \label{general_evol_eq_mq}
\ee
The UV-limit of the above equations can be worked out easily. Indeed, in this limit the function $g$ for $\rho=\rho_*$  is given by
\beq
e^{g}\approx {(R^2+\rho_*^2)^{{1\over 2b}}\over b}\ ,
\label{g(R)_UV}
\eeq
and we infer from (\ref{running_mass}) that the running mass is given by
\beq
m_q\,\approx\,{1\over 2\pi\,\alpha'^{3/2}\,b}\,\int_0^{R_0}\,
(R^2+\rho_*^2)^{{1\over 2b}-{1\over 2}}\,\approx\,{\Lambda^{1-b}\,R_0\over  2\pi\,\alpha'^{3/2}\,b}\,\,,
\label{mq_deep_UV}
\eeq
where, in the last step, we took the limit in which $\Lambda$ is large. Moreover, plugging (\ref{g(R)_UV}) into (\ref{general_evol_eq_mq}) we arrive at
\beq
\frac{\partial m_q}{\partial\log\Lambda} = m_q - \frac{1}{2\pi\alpha'^{3/2}} \frac{R_0}{(R_0^2+\Lambda^{2b})^{\frac{b-1}{2b}}}  \ .
\label{UV_evol_eq_mq}
\eeq
Furthermore, in the deep UV limit in which $\Lambda$ is large, we can neglect the $R_0^2$ in the denominator and using (\ref{mq_deep_UV}) we conclude that the second term in (\ref{UV_evol_eq_mq}) is just $-b\,m_q$.  We finally obtain the evolution equation for the quark mass $m_q$:
\be
 \frac{\partial m_q}{\partial\log\Lambda} = -\gamma_m m_q \ ,
 \label{CS_eq}
\ee
where we introduced the anomalous dimension of the quark mass
\be\label{eq:gammam}
 \gamma_m=b-1 \ .
\ee
The result (\ref{CS_eq}) is precisely the Callan-Symanzik equation for the effective mass which we were looking for. 
We see in this way that the gravitational background describes a renormalization group flow of the full theory. Notice that $\gamma_m\,=\,0$ for the unflavored case (and thus there is no running of $m_q$ when there are no dynamical flavors) and $\gamma_m\to 1/4$ as $\hat\epsilon\to\infty$.


\section{Entanglement entropy of strips}\label{sec:strip}
We want to characterize how information is distributed across scales in the wavefunction of our Chern-Simons theory with massive matter.  To this end, we can consider the entanglement of degrees of freedom within spatial regions $A$ with degrees of freedom in complementary regions $\bar{A}$.  The flow of the associated entanglement entropy with the size of $A$ characterizes the spatial organization of quantum information across scales.   Doing this computation directly within field theory is difficult, but is straightforward in the holographic dual. 

In the holographic presentation, where the Chern-Simons theory lives on the conformal boundary of spacetime, the entanglement entropy of a QFT region $A$ bounded by $\partial A$ is $1/4G_N$ times the area of a minimal surface $\Sigma$ anchored to $\partial A$ on the spacetime boundary \cite{Ryu}.  $\Sigma$ is thus an eight-dimensional surface embedded in the ten-dimensional bulk spacetime, and the entanglement entropy is
\beq
S_{A}\,=\,{1\over 4\,G_{10}}\,\int_{\Sigma}\,
d^8\xi\,e^{-2\phi}\,\sqrt{\det g_8}\,\,,
\label{entropy_functional}
\eeq
where the $\xi$'s are coordinates on $\Sigma$, $G_{10}$ is the ten-dimensional Newton constant ($G_{10}=8\pi^6$ in our units), $\phi$ is the dilation, and $g_8$ is the induced metric on $\Sigma$ in the string frame.   $\Sigma$ is a minimal surface, and hence minimizes the area integral on the right hand side.

We take our Chern-Simons theory to live on a plane $(x_1,x_2,t)$ and the region $A$ to be  $-{l\over 2}\,\le x^1\,\le +{l\over 2}$. This is a strip that is unbounded in $x_2$ and has a width $l$ in $x_1$.   To compute the entanglement entropy of $A$ we have to find the minimal surface $\Sigma$ in the geometry (\ref{eq:metric}) which is anchored on the edges of the strip at the spacetime boundary $x \to \infty$.  By symmetry, $\Sigma$ will lie in a constant time slice and be translation invariant in $x_2$ while also  filling up the compact directions.   Thus the embedding in ten-dimensional space will be specified by a single function $x=x(x^1)$ which describes the radial position of the surface at different values of $x_1$.    In terms of $x'= dx/dx^1$, we then find that the entanglement entropy (\ref{entropy_functional}) becomes
\beq
S(l)\,=\,{V_6 L_y\over 4 \,G_{10}}\,
\int_{-{l\over 2}}^{+{l\over 2}}\,dx^1\,\sqrt{H(x)}\,\sqrt{1+G(x)\,x'^{\,2}}\,\,,
\label{entropy_functional_strip}
\eeq
where  $V_6=32\,\pi^3/3$ is the volume of a unit $\mathbb{CP}^3$, $L_y=\int dx^2$ is (formally) the infinite length of the strip, and the functions  $G(x)$  and $H(x)$  are defined as:
\beq
G(x)\equiv {e^{2g}\,h\over x^2} \qquad ; \qquad H(x)\,\equiv\,h^2\,e^{-4\phi}\,e^{8f+4g}\,\,.
\label{G_H_def}
\eeq

The equation for the minimal surface embedding can be derived by minimizing $S(l)$ regarded as a functional of $x(x^1)$, and shows that 
\beq
x'\,=\,\pm {1\over \sqrt{G(x)}}\,\sqrt{{H(x)\over H_*}\,-\,1}\,\,,
\label{eq:xprime}
\eeq
where $H_*$ is the value of $H$ at the turning point $x=x_*(l)$ where the minimal surface reaches deepest into the bulk.
We can use this to covert integrals over $x^1$ into integrals over $x$ via $\int_{-l/2}^{l/2} dx^1 \to 2 \int_{x_*}^{\infty} dx/x' $.  Here the factor of 2 comes from symmetry around the turning point, and we have used the fact that the surface is anchored on the spacetime boundary at $x = \infty$.
Making this change of integration variable, which trades the strip width $l$ for the bulk turning point $x_*$, the entanglement entropy is
\beq
S(l)\,=\,{V_6 L_y\over 2 G_{10}}\,
\int_{x_*}^{\infty}\,dx\,{\sqrt{G(x)}\,H(x)\over \sqrt{H(x)-H_*}}\,\,.
\label{S_strip_total}
\eeq
In general $S(l)$ can be determined by  numerical integration of $G(x)$ and  $H(x)$  following \cite{Bea:2013jxa}, but we will see that an explicit analytic expansion can be developed when the number of flavors $N_f$ is much smaller than the number of colors $N$.

\subsection{$N_f \ll N$:  the few flavor limit}
The holographic entanglement entropy of strips can be determined analytically  in powers of $\hat{\epsilon} = (3/4) (N_f/N)\lambda$ using the few flavor expansion of the dual geometry in Appendix~\ref{sec:flavorexp}.   We will separately solve the cases when the turning point of the minimal surface is outside the cavity ($x_* > 1$) and when it is inside the cavity ($x_* < 1$).

\subsubsection{Turning point outside the cavity ($x_*\geq 1$)}
To compute the entanglement entropy we must evaluate the integral (\ref{S_strip_total}) using the definitions  (\ref{G_H_def}) of $G$ and $H$.  The $\hat\epsilon$ expansion in Appendix~\ref{sec:flavorexp} implies that
\begin{equation}
   G(x) = G_0 (x) ( 1 + \hat{\epsilon} G_1(x) ) + \mathcal{O}(\hat{\epsilon}^2) ~~;~~ H(x) = H_0 (x) ( 1 + \hat{\epsilon} H_1(x) ) + \mathcal{O}(\hat{\epsilon}^2) \label{eq:G_eps_expansion}  \ ,
\end{equation}
where
\begin{eqnarray}
  & G_0 = \frac{L_{ABJM}^4}{r_q^2} \frac{1}{x^4}  ~~;~~       H_0 = \frac{k^4 r_q^4 L_{ABJM}^4}{16} x^4 \label{eq:G_0} \\
  & G_1 = \frac{25-28 x^2-189 x^4+140 x^4 \log x}{280 x^4} ~~;~~    H_1 = - \frac{5+84 x^2-329 x^4+140 x^4 \log x}{140 x^4} 
\end{eqnarray}
outside the cavity, {\emph{i.e.}}, when $x > 1$. Here $L_{ABJM} = (2\pi^2 N/k)^{1/4}$ is the length scale of the AdS$_4$ part of the metric, and $k$ is the Chern-Simons level. 
The integral for $S(l)$ can then be expanded as
\begin{align}
   S(l) &= \frac{V_6 L_y}{2 G_{10}} \int_{x_*}^\infty dx \frac{ \sqrt{G_0(x) H_0(x)}}{\sqrt{1- \frac{H_0(x_*)}{H_0(x)}}} ( 1 + \hat{\epsilon} I_1(x) ) + \mathcal{O}(\hat{\epsilon}^2) \ ,
\end{align}
where
\begin{align}
   I_1(x) = \frac{ \left( 1- \frac{H_0(x_*)}{H_0(x)}\right) G_1(x) + \left(1-2\frac{H_0(x_*)}{H_0(x)}\right) H_1(x)+\frac{H_0(x_*)}{H_0(x)} H_1(x_*)}{2 \left( 1- \frac{H_0(x_*)}{H_0(x)} \right) } \ .
\end{align}
Entanglement entropy in quantum field theory has a UV divergence because of the large number of high frequency modes that straddle the boundary of the entanglement cut.  In the holographic computation this divergence reappears as an infinite contribution to the area of the corresponding minimal surface coming from the regions near the AdS boundary.  This divergence must be removed to identify the physically interesting finite parts.   There are divergent contributions to $S(l)$ coming from two  terms in the integrand:
\begin{equation}
    \frac{ \sqrt{G_0(x) H_0(x)}}{\sqrt{1- \frac{H_0(x_*)}{H(x)}}} \sim \frac{1}{4} k^2 L_{ABJM}^4 r_q ~~~~;~~~~
    I_1 (x) \sim \frac{67}{80} - \frac{\log x}{4} \ .
\end{equation}
The divergent integrals are as follows
\bea
 S^\text{div} = \lim_{\Lambda\to\infty}\frac{V_6 L_y}{2 G_{10}} \frac{k^2 r_q L_{ABJM}^4}{4}\left[\Lambda+\hat\epsilon \left(\frac{87}{80} \Lambda - \frac{\Lambda}{4} \log \Lambda\right) \right] \ .
\eea
To leading order in $\hat\epsilon$ the regularized entanglement entropy $S^\text{reg} = S-S^\text{div}$ obtained by subtracting the divergent contributions  from $S(l)$ can be written as
\begin{equation}
   S^\text{reg}(l) = S^{\text{reg},0}(l)+\hat\epsilon S^{\text{reg},\hat\epsilon}(l) 
\label{sregeqn}
\end{equation}
The flavorless  ($\hat\epsilon$-independent) part evaluates to the ABJM result
\begin{align}
   S^{\text{reg},0} &=  \frac{V_6 L_y}{2 G_{10}} \frac{k^2 r_q L_{ABJM}^4}{4} \left[ \int_{x_*}^\infty dx \left( \frac{1}{ \sqrt{1-(x_*/x)^4}} - 1 \right) - x_* \right]  \label{eq:S_xs_eps_integral1} \\
   & = - \frac{V_6 L_y}{2 G_{10}} \frac{k^ 2 r_q L_{ABJM}^4}{5} \frac{\sqrt{2} \pi^{3/2}}{\Gamma\left(\frac{1}{4}\right)^2} x_*
\end{align}
and the matter-dependent part evaluates to
\begin{align}
   S^{\text{reg},\hat\epsilon} = 
    \frac{V_6 L_y}{2 G_{10}} \frac{k^2 r_q L_{ABJM}^4}{4}  & \Bigg[ \int_{x_*}^\infty dx \left( \frac{I_1(x)}{ \sqrt{1-(x_*/x)^4}} - \left( \frac{67}{80} - \frac{\log x}{4} \right)  \right) 
    - \left( \frac{87}{80} x_* - \frac{x_*}{4} \log x_* \right) \Bigg] &
    \label{eq:S_xs_eps_integral2} \\  
  =  \frac{V_6 L_y}{2 G_{10}} \frac{k^ 2 r_q L_{ABJM}^4}{4} & \Bigg[ \frac{\pi^{3/2}}{56\sqrt{2}\Gamma\left(\frac{1}{4}\right)^2} \frac{1}{x_*^3} + \frac{12\pi^2-5\Gamma\left(\frac{1}{4}\right)^4}{40\sqrt{2\pi} \Gamma\left(\frac{1}{4}\right)^2} \frac{1}{x_*}   
  - \frac{87\pi^{3/2}}{40\sqrt{2}\Gamma\left(\frac{1}{4}\right)^2} x_* & \nonumber 
  \\
& + \frac{\pi^{3/2}}{2\sqrt{2}\Gamma\left(\frac{1}{4}\right)^2} x_* \log x_* \Bigg] \ .
\end{align}

Finally, we must revert the relation between strip width and turning point $l(x_*)$ to obtain $x_*(l)$.    This gives
\begin{equation}
   l = 2 \int_{x_*}^\infty dx \frac{ \sqrt{G(x)}}{ \sqrt{ \frac{H(x)}{H(x_*)}-1}}
   = \frac{A}{x_*} + \hat{\epsilon} \left( \frac{B}{x_*} + C \frac{\log x_*}{x_*} + D x_*^{-3} + E x_*^{-5} \right) \ ,
    \label{eq:l_xs_int} \\
\end{equation}
where
\begin{gather}
   A = \frac{2 L_{ABJM}^2}{r_q} \frac{\sqrt{2}\pi^{3/2}}{\Gamma\left(\frac{1}{4}\right)^2}, \qquad B = -\frac{2 L_{ABJM}^2}{r_q} \frac{7 \pi ^{3/2}}{40 \sqrt{2} \Gamma \left(\frac{1}{4}\right)^2}, \qquad C = \frac{2 L_{ABJM}^2}{r_q} \frac{\pi ^{3/2}}{2 \sqrt{2} \Gamma \left(\frac{1}{4}\right)^2} \ , \nonumber \\
   D = \frac{2 L_{ABJM}^2}{r_q} \frac{36 \pi ^2-5 \Gamma \left(\frac{1}{4}\right)^4}{120 \sqrt{2 \pi } \Gamma \left(\frac{1}{4}\right)^2}, \qquad E = \frac{2 L_{ABJM}^2}{r_q} \frac{\pi ^{3/2}}{56 \sqrt{2} \Gamma \left(\frac{1}{4}\right)^2} \ .
\end{gather}
To leading order, $x_* = A/l$. We can substitute this value for $x_*$ into the $\hat\epsilon$ terms in (\ref{eq:l_xs_int}) to get 
$l = \frac{A}{x_*} + \hat{\epsilon} \left( \frac{B}{A} l + \frac{C}{A} \log \left( \frac{A}{l} \right) l + \frac{D}{A^3} l^3 + \frac{E}{A^5} l^5 \right) + \mathcal{O}(\hat{\epsilon}^2)$.  Solving this expression for $x_*$ gives
\begin{equation}
x_* = \frac{A}{l-\hat{\epsilon}(\dots)} = \frac{A}{l} \left[ 1 + \hat{\epsilon} \left( \frac{B}{A} + \frac{C}{A} \log \left( \frac{A}{l} \right) + \frac{D}{A^3} l^2 + \frac{E}{A^5} l^4 \right)  \right] + \mathcal{O}(\hat{\epsilon}^2) \ .
\end{equation}

Substituting for $x_*(l)$ in (\ref{sregeqn}) and trading $L_{ABJM}$ (\ref{ABJM-AdSradius}), $G_{10}$ (\ref{entropy_functional}), and  $V_6$ (\ref{entropy_functional_strip}) for $\lambda$ (\ref{epsilonhatdef}) and $r_q$ (\ref{eq:rq}), we get the  regularized entanglement entropy as function of strip width $l$:
\begin{align}
   \frac{\lambda S^\text{reg}}{L_y N^2 r_q} = -\frac{4 \sqrt{2} \pi ^3 (1+\hat\epsilon)}{3 \Gamma \left(\frac{1}{4}\right)^4} \frac{\sqrt{\lambda}}{r_q l} -\frac{\hat\epsilon  \Gamma \left(\frac{1}{4}\right)^4}{144 \sqrt{2} \pi ^4} \frac{r_q l}{\sqrt{\lambda}} + \mathcal{O}(\epsilon^2)
   \ . \label{eq:EEoutside}
\end{align}
Note that all the logarithms have canceled out.  Since we were working with an $\hat\epsilon$ expansion of the metric that applies outside the cavity, (\ref{eq:EEoutside}) is valid if the turning point of the minimal surface is at $x_* \geq 1$.  This occurs so long as 
\begin{equation}
\frac{r_q l}{\sqrt{\lambda}} < \frac{r_q l_q}{\sqrt{\lambda}} = \frac{r_q l_\text{crit}}{\sqrt\lambda} - \frac{\sqrt{\pi } \left(7 \Gamma \left(\frac{1}{4}\right)^4-24 \pi ^2\right)}{84 \Gamma \left(\frac{1}{4}\right)^2}\hat{\epsilon} ~~~;~~~ 
  \frac{r_q l_\text{crit}}{\sqrt\lambda} = \frac{4\pi^{5/2}}{\Gamma\left(\frac{1}{4}\right)^2}   \ ,  \label{eq:lcrit}
  \end{equation}
where $l_q = l(x_*=1)$ from (\ref{eq:l_xs_int}) is the width of the strip just touching the cavity.  In the following we will be expanding most quantities to first order in $\hat\epsilon$, and thus in some expressions the $\hat\epsilon$ term in $l_q$ can be ignored.  In these cases  $l_{\rm crit}$ will effectively be the width of the strip whose bulk minimal surface just touches the cavity in the gravity solution.

\subsubsection{Turning point inside the cavity ($x_*<1$)}
When the extremal surface penetrates the cavity ($x_*<1$), the entanglement entropy is still given by (\ref{sregeqn}) with $G$, $H$, $G_0$, and $H_0$ given by (\ref{eq:G_eps_expansion}),(\ref{eq:G_0}) but now
\begin{equation}
   G_1 = -\frac{4}{5}+\frac{4}{35} x^3 ~~~~;~~~~
   H_1 = \frac{8}{5} + \frac{4}{35} x^3 \ .
\end{equation}
Integrals must now be separately evaluated inside and outside the cavity using the IR and UV parts of the solution discussed in the previous section.
Regularization works in the same way as previously because divergences always come from the region near the AdS boundary, and the $\hat\epsilon$-independent terms are unchanged because these terms represent the geometry in the absence of flavors in the field theory.   The flavor correction when  $0<x_*<1$ evaluates to
\begin{align}
   S^{\text{reg},\hat\epsilon} &= \frac{V_6 L_y}{2 G_{10}} \frac{k^2 r_q L_{ABJM}^4}{4}  \Bigg[ \frac{1}{10} \left(-9 (\re(K(2)) -2 K(-1)+E(-1))+5\mathcal{E}(x_*)-5 \mathcal{F}(x_*)\right) x_* \nonumber \\
   &-\frac{\re(K(2))+\mathcal{F}(x_*)}{2 x_*} -\frac{1}{35} (\re(K(2))-2 K(-1)+E(-1)) x_*^4\Bigg], \label{eq:S_xs_in}
\end{align}
where $K(m)$ and $E(m)$ are the complete elliptic integrals of first and second kinds, respectively, and
\begin{align}
   \mathcal{E}(x_*) &= \im\left(E\left(\left.\csc ^{-1}\left(x_*\right)\right|-1\right)\right) \\
   \mathcal{F}(x_*) &= \im\left(F\left(\left.\csc ^{-1}\left(x_*\right)\right|-1\right)\right),
\end{align}
where $F(\phi | m)$ and $E(\phi | m)$ are the elliptic integrals of first and second kinds, respectively, which in our conventions are defined by
\begin{align}
   F(\phi|m) &= \int_0^\phi \frac{1}{\sqrt{1-m \sin^2\theta}} d\theta \\
   E(\phi|m) &= \int_0^\phi \sqrt{1-m \sin^2\theta} d\theta \ .
\end{align}
The complete elliptic integrals are then $K(m)=F(\frac{\pi}{2}|m)$ and $E(m)=E(\frac{\pi}{2}|m)$.


Next we must express the turning point $x_*$ in terms of the strip width in the field theory. To do this we first compute $l(x_*)$ for $0<x_*<1$ up to $\mathcal{O}(\hat\epsilon^2)$ by evaluating (\ref{eq:l_xs_int}) in two parts, first inside and then outside the cavity. The result is
\begin{align}
   l &= \frac{2 L_{ABJM}^2}{r_q} \Bigg[ \frac{\sqrt{2}\pi^{3/2}}{\Gamma\left(\frac{1}{4}\right)^2} \frac{1}{x_*} + \hat{\epsilon} \Bigg( \frac{1}{35} (-\re(K(2))+2 K(-1)-E(-1)) x_*^2 -\frac{\sqrt{1-x_*^4}}{3 x_*^2}\nonumber \\
   & + \frac{-\re(K(2))-\mathcal{F}\left(x_*\right)}{6 x_*^3} + \frac{\re(K(2))-2 K(-1)-5 \mathcal{E}\left(x_*\right)+5 \mathcal{F}\left(x_*\right)+E(-1)}{10 x_*} \Bigg) \Bigg]. \label{eq:l_xs_in}
\end{align}
Our final expression for the entanglement entropy in this regime is given by expressing (\ref{eq:S_xs_in}) in terms of $l$ which can be achieved by reverting (\ref{eq:l_xs_in}). The result is
\begin{align}
   \frac{\lambda S^\text{reg}}{L_y N^2 r_q} &= -\frac{4 \sqrt{2} \pi ^3}{3 \Gamma \left(\frac{1}{4}\right)^4} \frac{\sqrt{\lambda}}{r_q l} + \frac{\hat{\epsilon}}{9\pi} \Bigg[\sqrt{1-\frac{l_\text{crit}^4}{l^4}} -\left( \mathcal{F}\left(\frac{l_\text{crit}}{l}\right) + \re(K(2))\right) \frac{l}{l_\text{crit}} + \nonumber \\
   & + \left(\frac{3}{4 \sqrt{2 \pi }}\frac{\Gamma \left(\frac{1}{4}\right)^4-8 \pi ^2}{\Gamma
   \left(\frac{1}{4}\right)^2}+3
   \left(\mathcal{E}\left(\frac{l_{\text{crit}}}{l}\right)-\mathcal{F}\left(\frac{l_{\text{crit}}}{
      l}\right)-\re(K(2))\right) \right) \frac{l_\text{crit}}{l} \Bigg] \nonumber \\
   & + \mathcal{O}(\hat\epsilon^2), \quad\quad 
    l > l_\text{crit}  \ . \label{eq:EEinside}
\end{align}

\begin{figure}[!ht]
   \centering
   \includegraphics[width=0.7\textwidth]{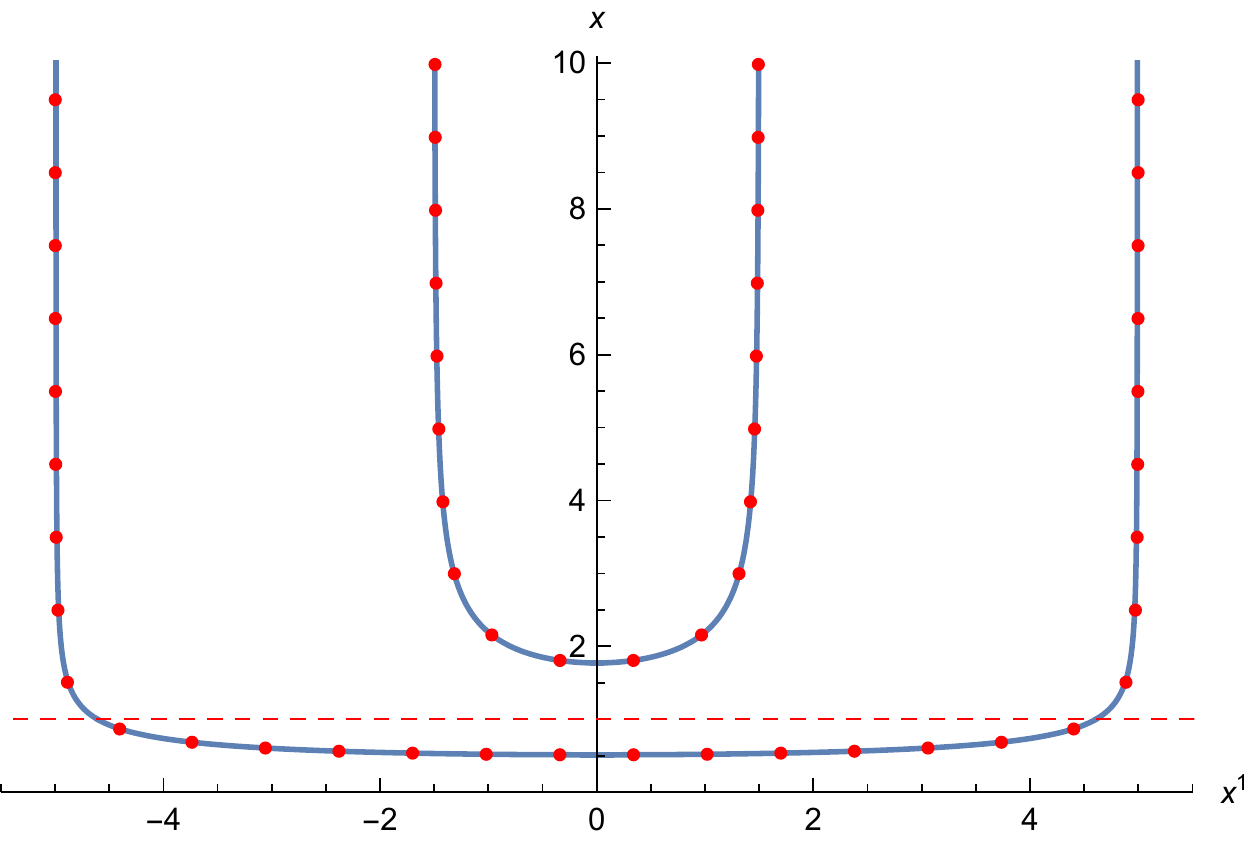}
   \caption{Minimal surfaces for $\frac{r_q l}{\sqrt\lambda}= 3, 10$ with $\hat\epsilon=(3/4)(N_f/N)\lambda = 0.1$. The minimal surface for the wider boundary strip penetrates the cavity (location marked by horizontal dashed line).  Analytical results for the surfaces (curves) match numerical solutions (dots) of the equation of motion (\ref{eq:xprime})  extremely well.}
   \label{fig:embeddings}
\end{figure}

\begin{figure}[!ht]
   \centering
   \includegraphics[width=0.7\textwidth]{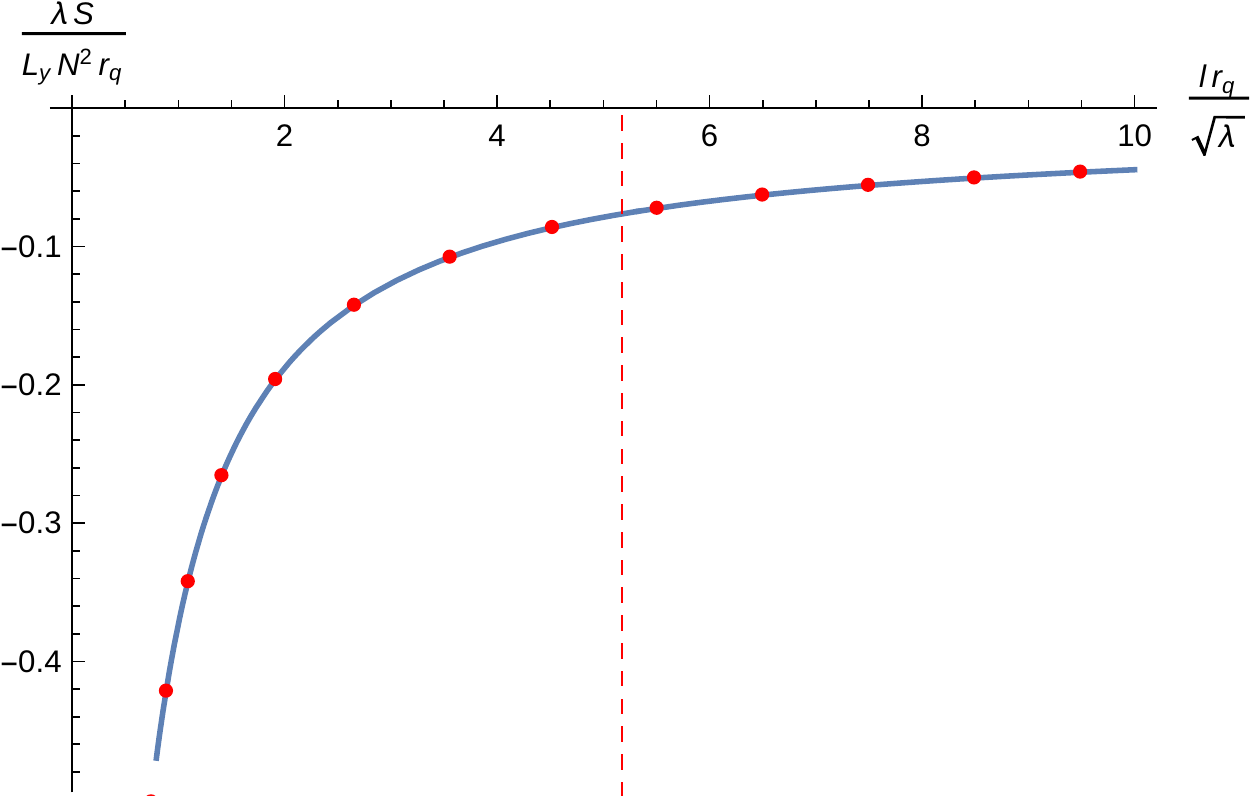}
   \caption{Regularized entanglement entropy as a function of strip width for $\hat{\epsilon}= (3/4)(N_f/N)\lambda = 0.1$.  The numerical (points) and analytical (curve) results match perfectly. The vertical dashed line is the location of the cavity.}
   \label{fig:hee}
\end{figure}

Fig.~\ref{fig:embeddings} shows the embeddings and Fig.~\ref{fig:hee}  the regularized entanglement entropy for a single strip in the few flavor limit.   The analytical result for narrow strips  (\ref{eq:EEoutside}) and for wide strips  (\ref{eq:EEinside})  both match the numerical results extremely well.

\subsection{Dimension of the RG driving operator}
If we were studying the ABJM theory with massless quarks, there would have been no RG flow because the theory would have been conformal.  However, we are working in a theory in which the quark constituent mass has been deformed from $m=0$ to $m' = 0 + \delta m$ by the addition of the operator $\bar\psi\psi$ in the Hamiltonian.  This is a relevant deformation -- the mass runs to $0$ in the UV and to a finite value in the IR.     From the bulk AdS$_4$ perspective this deformation corresponds to turning on a scalar field.   In the top-down ten-dimensional embedding that we are analyzing, there are no scalars, as these only appear after the reduction to 4 dimensions; in 10d language the excited 4d scalar of interest is visible in non-trivial components of the metric.  In this section we will use the gravitational solution to work out the dimension of the relevant operator driving the renormalization group flow.   This dimension can be read off directly from the bulk metric, and also from the UV-behavior of the entanglement entropy, providing a cross-check on the formulas derived in the last section.\footnote{Our treatment partly follows \cite{Klebanov:2012yf}.}

Consider the holographic dual of any three-dimensional field theory near a UV fixed point. The four-dimensional part of the bulk metric must asymptote to $AdS_4$ and can be written near the boundary as
\begin{align}
   ds_4^2 = \frac{L_{UV}^2}{z^2} \left( \frac{dz^2}{f(z)} - dt^2 + dx^2 + dy^2 \right) \ , \label{eq:uv_metric}
\end{align}
where $f(z)=1+\mathcal{O}(z^\alpha)$ and the boundary lies at $z=0$. Departures from $AdS_4$, and thus the effects of RG flow, are encoded in the subleading behavior of $f(z)$. We now assume that the RG flow is driven by a relevant scalar perturbation $\mathcal{O}$ with a dimension $1/2<\Delta<3$. This is dual holographically to a scalar field $\phi$ with mass $m^2=\Delta(\Delta-3)$. The bulk action will then be
\begin{align}
   S_4 = \frac{1}{16\pi G_N^{(4)}} \int d^4 x \sqrt{-g_4} \left( R + \frac{6}{L_{UV}^2} - \frac{1}{2} \partial^\mu \phi \partial_\mu \phi - \frac{1}{2} m^2 \phi^2 + \ldots \right) \ .
\end{align}
When $\phi(z,\vec{x})=0$ the corresponding background solution reduces to $AdS_4$. The equation of motion for the scalar gives the near boundary behavior
\begin{align}
   \phi(z,\vec{x})=z^{3-\Delta} (\phi_0(\vec{x}) + \mathcal{O}(z^2)) \ 
\end{align}
and corresponds to a perturbation of the UV CFT by
\begin{align}
   S_3 = S_3^{(UV)} + \int d^3 x \,  \phi_0(\vec{x}) \, \mathcal{O}(\vec{x}) \ .
\end{align}
Let us focus on a uniform perturbation $\phi_0(\vec{x})=\text{const.}=\phi_0$. In this case the Einstein equations give the backreaction of the scalar field on the metric
\begin{align}
   f(z) = 1 + \frac{3-\Delta}{4} \phi_0^2 z^{2(3-\Delta)} \ .
\end{align}
Given the metric functions near the UV this can be used to find the operator dimension $\Delta$. 

The effect of RG flow is also encoded in the UV behavior of the entanglement entropy.  Consider the entanglement entropy of a strip in a theory with a gravity dual  (\ref{eq:uv_metric}).   For narrow strips the holographic entanglement entropy in the Einstein frame is
\begin{align}\label{eq:HEEatUV}
   S = \frac{1}{4 G_N^{(4)}} \int \sqrt{g_2} \ ,
\end{align}
where
\begin{align}
   ds_2^2 = \frac{L_{UV}^2}{z^2} \left( \left( \frac{z'^2}{f(z)} + 1 \right) (dx^1)^2 + (dx^2)^2 \right) 
\end{align}
is the induced metric on a bulk surface anchored to the strip and the prime denotes derivation with respect to $x^1$. Explicitly, this gives
\begin{align}
   S = \frac{L_y L_{UV}^2}{2 G_N^{(4)}} \int_0^{l/2} \frac{dx}{z^2} \sqrt{ \frac{z'^2}{f(z)}+1} \ ,
\end{align}
where $L_y$ and $l$ are the length and the width of the strip, respectively. To find the minimal surface we evaluate the Euler-Lagrange equations for $z(x)$, $   z' = \frac{ \sqrt{f(z)} \sqrt{z_*^4-z^4}}{z^2} $,
which, when inserted in (\ref{eq:HEEatUV}), yields
\begin{align}
   S &= \frac{L_y L_{UV}^2}{2 G_N^{(4)}} \int_0^{z_*} \frac{z_*^2 dz}{z^2 \sqrt{(z_*^4-z^4) f(z)}} \\
   &= \frac{L_y L_{UV}^2}{2 G_N^{(4)}} \int_0^{z_*} \left( \frac{z_*^2}{z^2 \sqrt{z_*^4-z^4}} - \frac{z_*^2 z^{4-2\Delta}}{2 \sqrt{z_*^4-z^4}} \frac{3-\Delta}{4} \phi_0^2 + \ldots \right) dz \ .
\end{align}
In addition to the usual divergence of areas of surfaces anchored to asymptotically AdS boundaries, there is a divergence coming from the scalar field-dependent term when $5/2\leq\Delta<3$.   Subtracting these divergences,
\be
 S^\text{div}=\lim_{\epsilon\to 0}\frac{L_y L_{UV}^2}{2 G_N^{(4)}} \left( \frac{1}{\epsilon} + \frac{\epsilon^{5-2\Delta}}{2(5-\Delta)} \frac{3-\Delta}{4} \phi_0^2 \right) \ ,
\ee
the regulated entanglement entropy $S^\text{reg}=S-S^\text{div}$ reads
\begin{align}
   S^\text{reg} = \frac{L_y L_{UV}^2}{2 G_N^{(4)}} \left( - \frac{\sqrt{2}\pi^{3/2}}{\Gamma\left(\frac{1}{4}\right)^2} \frac{1}{z_*} - \frac{\sqrt{\pi} \Gamma\left(\frac{5}{4}-\frac{\Delta}{2}\right)}{8\Gamma\left(\frac{7}{4}-\frac{\Delta}{2}\right)} \frac{3-\Delta}{4} \phi_0^2 z_*^{5-2\Delta} + \ldots \right) \ .
\end{align}

To work out the turning point as a function of strip width $z_*(l)$, we evaluate
\begin{align}
   l &= 2 \int_0^{z_*} \frac{z^2 dz}{\sqrt{f(z)}\sqrt{z_*^4-z^4}} = 2 \int_0^{z_*} \left( \frac{z^2}{\sqrt{z_*^4-z^4}} - \frac{z^{2+2(3-\Delta)}}{\sqrt{z_*^4-z^4}} + \ldots \right) dz \\
    &= \frac{2\sqrt{2}\pi^{3/2}}{\Gamma\left(\frac{1}{4}\right)^2} z_* - \frac{\sqrt{\pi} \Gamma\left(\frac{9}{4}-\frac{\Delta}{2}\right)}{4 \Gamma\left(\frac{11}{4}-\frac{\Delta}{2}\right)} \frac{3-\Delta}{4}\phi_0^2 z_*^{7-2\Delta} + \ldots \ .
\end{align}
This relation can be reverted to give
\begin{align}
   z_* = \frac{\Gamma\left(\frac{1}{4}\right)^2}{2\sqrt{2}\pi^{3/2}} l + \frac{2^{-14+3\Delta}\pi^{-23/2+3\Delta}\Gamma\left(\frac{1}{4}\right)^{16-4\Delta}\Gamma\left(\frac{9}{4}-\frac{\Delta}{2}\right)}{\Gamma\left(\frac{11}{4}-\frac{\Delta}{2}\right)} \frac{3-\Delta}{4} \phi_0^2 l^{7-2\Delta} + \ldots \ .
\end{align}
Finally, the entanglement entropy is
\begin{align}
   S^\text{reg} = \frac{L_y L_{UV}^2}{2 G_N^{(4)}} \left( - \frac{4\pi^3}{\Gamma\left(\frac{1}{4}\right)} \frac{1}{l} - \frac{2^{-32/2+3\Delta}\pi^{-7+3\Delta}\Gamma\left(\frac{1}{4}\right)^{10-4\Delta}\Gamma\left(\frac{5}{4}-\frac{\Delta}{2}\right)}{\Gamma\left(\frac{11}{4}-\frac{\Delta}{2}\right)} \frac{3-\Delta}{4}\phi_0^2 l^{5-2\Delta} + \ldots \right) \ . \label{eq:EEfinal}
\end{align}

In our massive ABJM setting we can identify
\begin{align}
   f(z)=\frac{L_0^2 z^2}{\sqrt{h(z)}} = 1 - \frac{\kappa^{2b}}{2} \left( h_2 - \frac{4 g_2}{2b-1} \right) (r_q z)^{2b} + \ldots
\end{align}
leading to
\begin{align}\label{eq:Deltavsb}
   \Delta = 3-b = 2-\gamma_m \ ,
\end{align}
for  the dimension of the leading operator driving the RG flow.  Here we used the expression for the mass anomalous dimension (\ref{eq:gammam}) in the last step.   We can likewise use (\ref{eq:EEfinal}) to derive the RG-driving operator dimension from  the UV-behavior of the entanglement entropy. It turns out that the leading correction to the entanglement entropy behaves as $\delta S^\text{reg}_\text{ABJM} \sim l^{-1+2b}$  (details in  Appendix \ref{app:UVexpansion}) which correctly implies (\ref{eq:Deltavsb}), $\Delta=3-b$.  This confirms the validity of our analytical expressions for the entanglement entropy.


\section{The flow of mutual information}\label{sec:mutual}

The mutual information between two entangling regions $A$ and $B$ is defined as
\begin{align}
   I(A,B) = S(A) + S(B) - S(A\cup B)  \  , \label{def:mutu}
\end{align}
and characterizes the information that is shared between these domains.  The flow of mutual information with the size of the entangling regions characterizes how information is shared across space at different scales.   We will characterize this flow for the mutual information between strips in four ways: (a) in terms of a ``phase transition'' defined by the separation distance between strips of fixed width at which the mutual information vanishes, (b) in terms of a c-function obtained from the mutual information that counts the degrees of freedom that are active at the scale of the strips, (c) in terms of a measure of ``extensivity'' that characterizes the additivity of the mutual information between a region $A$ and two other regions $B$ and $C$, and (d) in terms of mutual information between regions $A$ and $B$ conditioned on knowledge of a third region $C$.

\subsection{Mutual information transitions}
\label{sec:mutinfotrans}
\begin{figure}
   \centering
   \includegraphics[width=\textwidth]{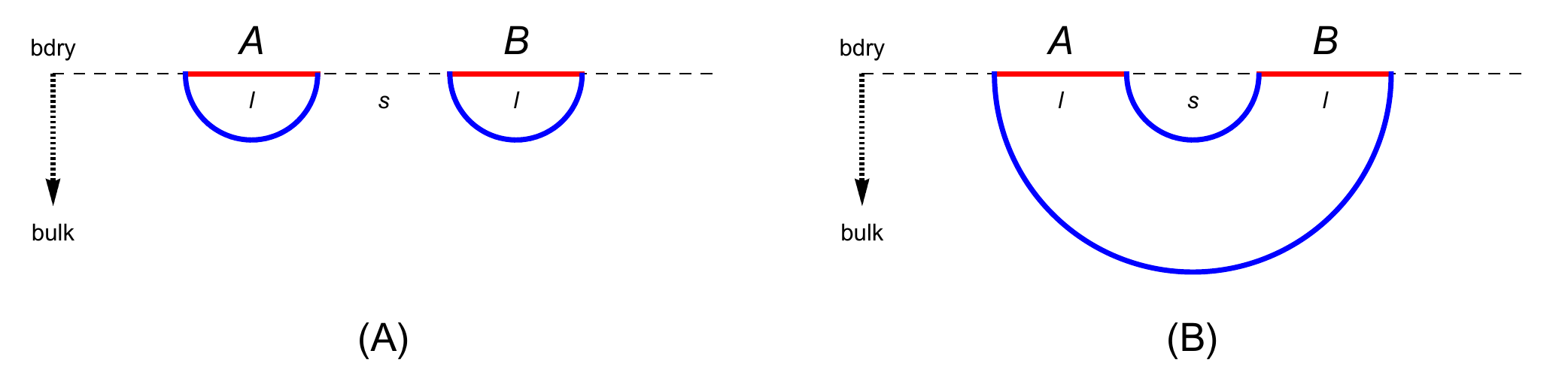}
   \caption{Bulk extremal surfaces (blue) that can contribute to the holographic mutual information between boundary strips $A$ and $B$ of length $l$.   ({\bf{A}}): When the separation $s$ of the strips is large the minimal area surface bounded by $A\cup B$ is the union of the minimal surfaces for $A$ and $B$ separately.  Thus the mutual information (\ref{def:mutu}) vanishes because $S(A) + S(B) = S(A\cup B)$.  ({\bf{B}}): When the separation $s$ of the strips is small, the minimal area surface bounded by $A\cup B$ is the union the pictured surfaces.   Thus the mutual information (\ref{def:mutu}) is non-vanishing because  $S(A) + S(B) > S(A\cup B)$.   If the theory is conformal (either because there are no quarks, or when they are massless) the text shows that this mutual information phase transition occurs when $s/l|_{\hat\epsilon=0}=(\sqrt{5}-1)/2 = 1/\varphi$ \cite{Alishahiha:2014jxa,Ben-Ami:2014gsa} where $\varphi$ is the golden ratio $(\sqrt{5} + 1)/2$.}
   \label{fig:strip_configurations}
\end{figure}

In theories with a holographic dual there is typically a scale-dependent phase transition in the mutual information:  $I$ is finite when two regions of fixed sizes are sufficiently close, but vanishes when the separation grows to a certain critical size.  Holographically, this transition occurs when the minimal bulk surface that approaches $A\cup B$ on the boundary fragments into the union of minimal surfaces for $A$ and $B$ (see Fig.~\ref{fig:strip_configurations}).  In the few flavor limit we will find an analytic expression for  the critical separation at which the phase transition occurs in the mutual information between strips.  The critical separation will be a function of the scale $l$ defined by the strips themselves.

To this end, let us consider a system of two parallel strips of width $l$ separated by a distance $s$. The transition between the two configurations in Fig.~\ref{fig:strip_configurations} happens when their entropies are equal: 
\begin{align}
   2 S^\text{reg}(l) = S^\text{reg}(2l+s) + S^\text{reg}(s) \label{eq:tpoint_eq_eps} \ .
\end{align}
Substituting the $\hat\epsilon$ expansion $S^\text{reg} = S^{\text{reg},0} + \hat{\epsilon} S^{\text{reg},\hat{\epsilon}}$ in (\ref{eq:tpoint_eq_eps}) we find
\begin{align}
   &2(S^{\text{reg},0}(l)+\hat{\epsilon} S^{\text{reg},\hat{\epsilon}}(l)) \nonumber \\
   & = S^{\text{reg},0}(2l+s) + \hat{\epsilon} S^{\text{reg},\hat{\epsilon}}(2l+l/\varphi) + S^{\text{reg},0}(s) + \hat{\epsilon} S^{\text{reg},\hat{\epsilon}}(l/\varphi)  \ ,
\end{align}
where we have introduced the golden ratio $\varphi=(\sqrt 5+1)/2$:
\be
 \frac{1}{\varphi} = \frac{\sqrt 5-1}{2} \ .
\ee
Solving for $s(l)$ gives
\be
\frac{s}{l}  =  \frac{1}{\varphi} + \hat{\epsilon} \frac{3 \left(\varphi+2\right) \Gamma \left(\frac{1}{4}\right)^4}{40 \sqrt{2} \pi ^3} \frac{\lambda^2 l}{N^2 L_y} \Big( 2 S^{\text{reg},\hat{\epsilon}}(l) -S^{\text{reg},\hat{\epsilon}}(2l+l/\varphi) - S^{\text{reg},\hat{\epsilon}}(l/\varphi) \Big) + \mathcal{O}(\epsilon^2) \ ,\label{eq:soverlanalytic}
\ee
where the entropies of individual configurations are given either by (\ref{eq:EEoutside}) or by (\ref{eq:EEinside}) depending on whether their lengths are smaller or larger than $l_{\text{crit}}$ 
given in (\ref{eq:lcrit}).  Close to the CFT fixed points in the UV and IR we find that 
\begin{align}
  {\text{UV}} \ : \ \frac{s}{l} &\approx \frac{1}{\varphi} + \hat{\epsilon}\frac{\Gamma\left(\frac{1}{4}\right)^8}{384(2\varphi-1)\pi^7} \left( \frac{r_q l}{\sqrt{\lambda}} \right)^2 \ , & 2l+l/\varphi < l_{\text{crit}} \\ 
   {\text{IR}} \ : \ \frac{s}{l} &\approx \frac{1}{\varphi} + \hat\epsilon \frac{32\sqrt 2 (5/2-\varphi)\pi^6}{15\Gamma\left(\frac{1}{4}\right)^4} \left( \frac{r_q l}{\sqrt{\lambda}} \right)^{-3} \ , & \frac{r_q l}{\sqrt{\lambda}}\to\infty \ .
\end{align}
In the UV limit the bulk minimal surfaces remain in the region $x > x_* = 1$, while in the IR limit all the surfaces penetrate to $x < x_*$.

\begin{figure}[!ht]
   \centering
   \includegraphics[width=0.7\textwidth]{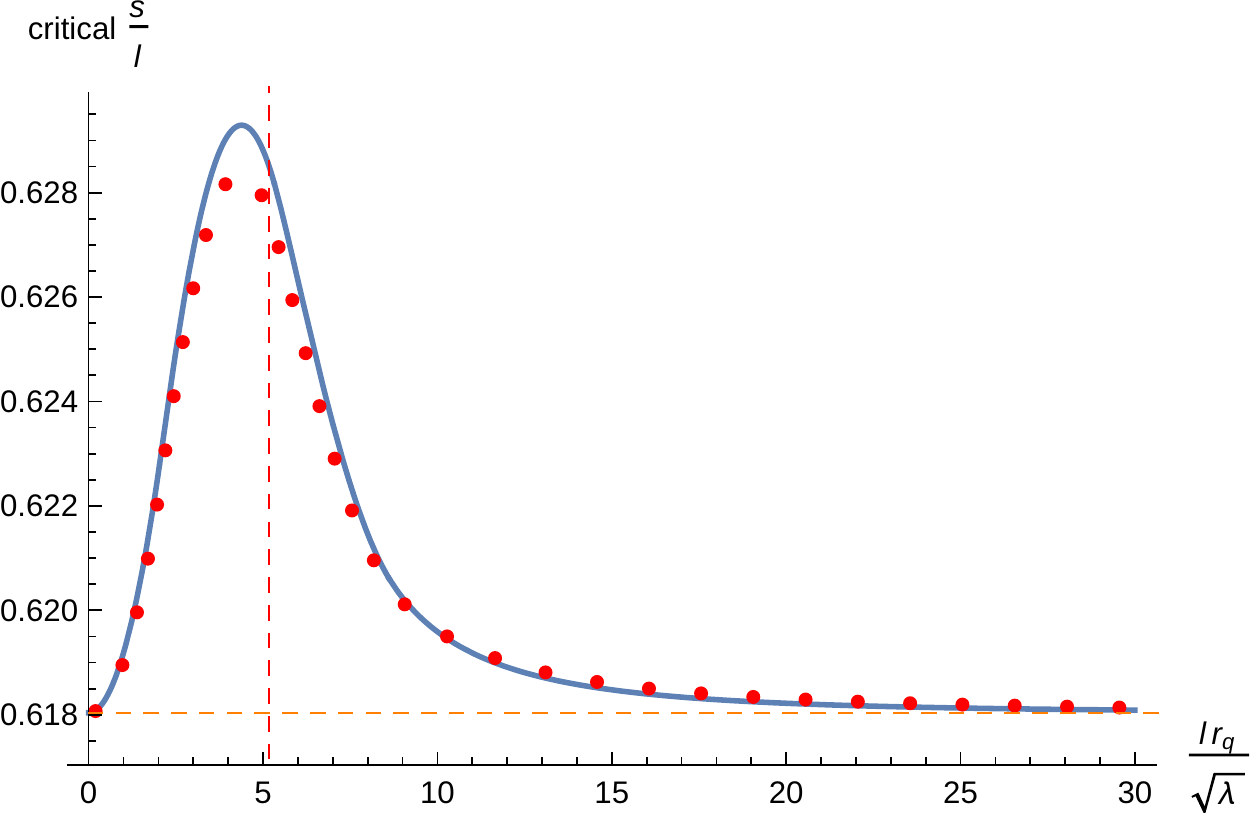}
   \caption{Critical values of  $s/l$ for $\hat\epsilon=0.1$,  at which the mutual information between strips of width $l$ separated by a spacing $s$ drops to zero.  The dashed horizontal orange line denotes the critical value in a CFT.   The numerical results (red dots)  closely match the analytical results (blue curve; Eq.~(\ref{eq:soverlanalytic})). Higher order in $\hat\epsilon$ corrections to (\ref{eq:soverlanalytic}) should produce an even closer match.     The vertical red dashed line is $l_q r_q/\sqrt{\lambda}$ where $l_q$ is the width of the strip whose bulk minimal surface just touches the cavity.}
   \label{fig:transition_points}
\end{figure}

To leading order in $\hat\epsilon$, $s(l)$ coincides  with the result for a conformal theory dual to AdS$_4$: $s(l)|_{\hat\epsilon=0} / l = 1/\varphi =(\sqrt{5}-1)/2 \approx 0.618$ \cite{Alishahiha:2014jxa,Ben-Ami:2014gsa}.\footnote{Interestingly, for the symmetric case of $m$ equally separated strips of equal width $l$, as also studied in \cite{Alishahiha:2014jxa,Ben-Ami:2014gsa}, the critical separation $l/s =\frac{\left(\frac{2}{m}\right)+\sqrt{\left(\frac{2}{m}\right)^2+4}}{2}$ is the so-called ``metallic mean''.}    It also tends to this value in the conformal limits well above and below the mass scale of the quarks ($l \to 0$ and $l \to \infty$).   At intermediate scales $l$, the mutual information transition occurs at a separation $s$ between strips that is greater than in a conformal field theory (Fig.~\ref{fig:transition_points}), and is maximized at a scale where the bulk minimal surface almost reaches the ``cavity'' associated to the quark mass scale.    Interestingly, and surprisingly, this suggests that information is more non-locally shared away from the conformal fixed points.  

The non-monotonic behavior in Fig.~\ref{fig:transition_points} also implies that in a range of  fixed values of $s/l$ (between the peak and the horizontal dashed line), the mutual information between strips will vanish both for large $l$ and for small $l$.  In both these cases the massive ABJM theory is in a phase where the entropy of the union of strips is given holographically by the disconnected surfaces in Fig.~\ref{fig:strip_configurations}A.   For these $s/l$ there is also an intermediate range of $l$ (determined by the intersection of a horizontal line of fixed $s/l$ with the critical curve in Fig.~\ref{fig:transition_points}) in which the mutual information is non-zero, or equivalently, in which the entropy of the union of strips is given by the connected surfaces in Fig.~\ref{fig:strip_configurations}B.  This has a remarkable implication about the organization of quantum information in this theory.  Apparently, two strips at a fixed $s/l$ that share mutual information can be mutually disentangled by {\it increasing} the strip width at fixed $s/l$.  This sort of purification by expansion does not happen in a CFT where strips with a given $s/l$ will always either share some mutual information or none at all, independently of the value of $l$.

\begin{figure}[!ht]
   \centering
   \includegraphics[width=1.\textwidth]{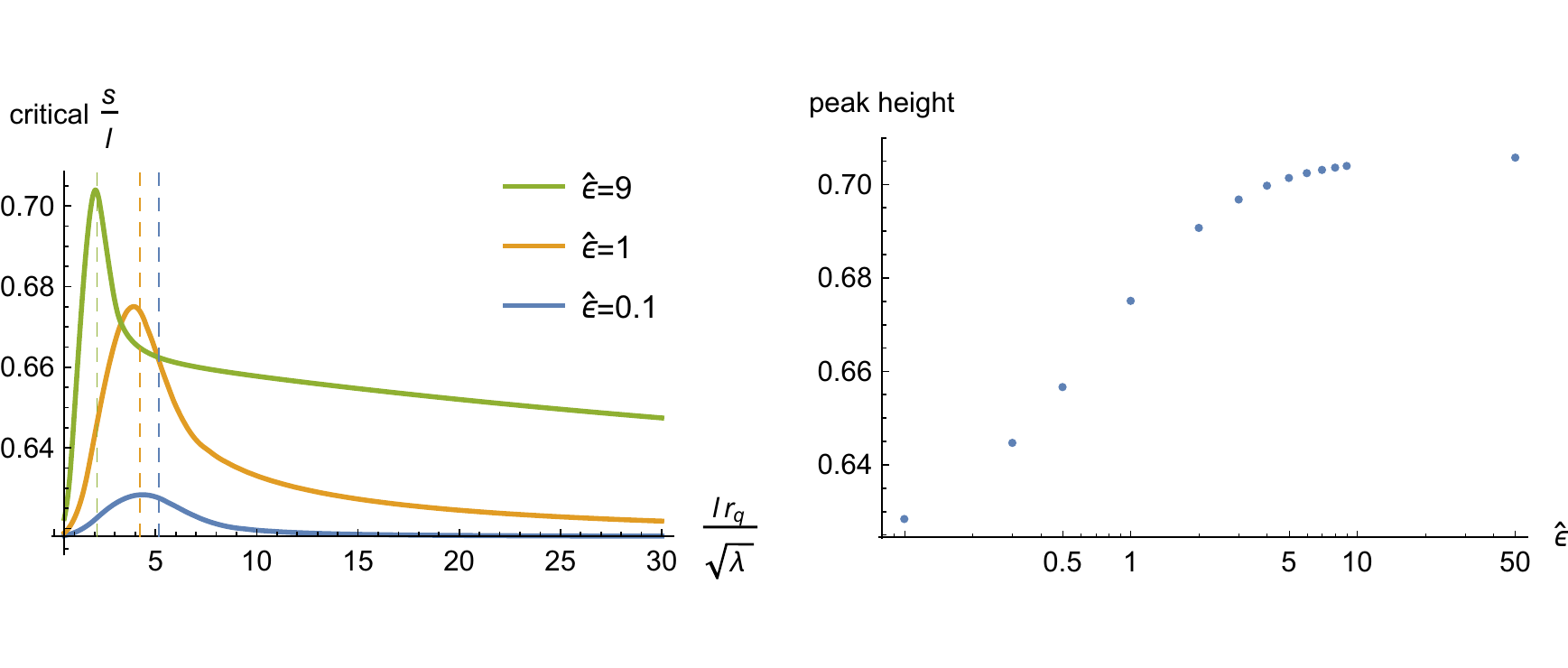}\hfil
      ({\bf A}) \hfil   \hspace{1.5in}  ({\bf B})  \hfil 
   \caption{({\bf{A}}) Numerical analysis of the critical value of $s/l$ ($s =$ strip separation, $l =$ strip width) at which the mutual information transition occurs for varying numbers of flavors: $(3/4)(N_f/N)\lambda = \hat\epsilon=0.1,1,9$.  As $N_f$ increases, the peak deviation from the conformal value of $s/l$ grows, and occurs further in the UV (smaller strip widths $l$).  The vertical dashed lines are  $l_q r_q/\sqrt{\lambda}$ where $l_q$ is the width of the strip whose bulk minimal surface just touches the cavity for each $\hat\epsilon$. ({\bf{B}}) Peak deviation of the critical $s/l$ as a function of the number of flavors.  The numerical evidence confirms that in the absence of flavors ($\hat\epsilon \to 0$) we return to the conformal result $s/l = 1/\varphi$ where $\varphi$ is the golden ratio, and suggests that in the limit of many flavors ($\hat\epsilon \to \infty$) the deviation from the conformal value saturates. }
   \label{fig:peak_height}
\end{figure}

\begin{figure}[!ht]
   \centering
\includegraphics[width=\textwidth]{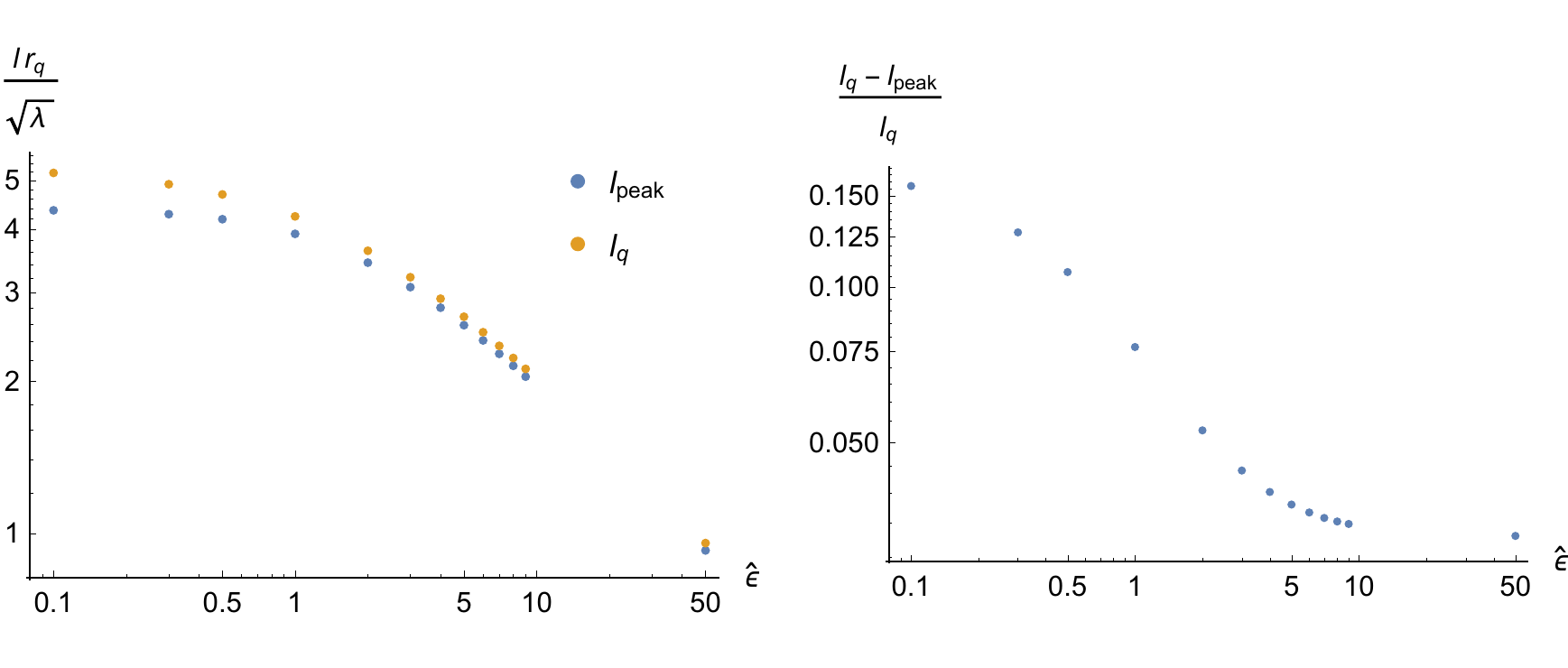}
 \\
      ({\bf A}) \hfil   \hspace{1.5in}  ({\bf B})  \hfil 
   \caption{Numerical analysis of the scale $l_{peak}$ at which the critical value of $s/l$ maximally deviates from the conformal value.  ({\bf{A}}) Comparison of $l_{peak} r_q/\sqrt{\lambda}$ with $l_q r_q/\sqrt{\lambda}$ as a function of the number of flavors ($\hat\epsilon = (3/4)(N_f/N)\lambda$).  Here $l_q$ is the strip width at which the corresponding bulk minimal surface just touches the cavity; i.e., this is the scale set by the quark masses.  The absolute difference between these scales shrinks in the many flavor limit.  ({\bf{B}})  The relative difference between these scales also shrinks in the many flavor limit, but may be approaching a finite value.
   }
   \label{fig:peak_position}
\end{figure}

As the number of quarks or, equivalently, $\hat\epsilon = (3/4)(N_f/N)\lambda$, increases, the peak in Fig.~\ref{fig:transition_points} becomes higher (Fig.~\ref{fig:peak_height}) and narrows but also acquires a longer tail towards the infrared (large $l$).  The tail means that the deviation from the conformal theory persists deeper into the infrared.   When the $\hat\epsilon > 1$, the transition occurs at an $s/l$ far from the conformal value even at scales well into the infrared.   Of course in the deep infrared limit the theory approaches a conformal fixed point (an AdS$_4$ geometry in the bulk description) and so the transition $s/l$ eventually returns to the conformal value.  Meanwhile the scale $l$ at which there is the  greatest deviation from the conformal result ({\emph{i.e.}}, the peak of the bump in Fig~\ref{fig:transition_points}) moves further into the UV as $N_f$ increases (Fig.~\ref{fig:peak_position}).  At the same time this scale approaches more closely the strip width $l_q$ for which the corresponding minimal surface just touches the cavity.  It would be interesting to study the $\hat\epsilon\to\infty$ limit to see whether the peak height approaches a finite value  as suggested by the numerical analysis in Fig.~\ref{fig:peak_height} and whether the peak location coincides precisely with $l_q$.

\subsection{c-functions: flow of the number of degrees of freedom}
\begin{figure}[!ht]
   \centering
   \includegraphics[width=0.7\textwidth]{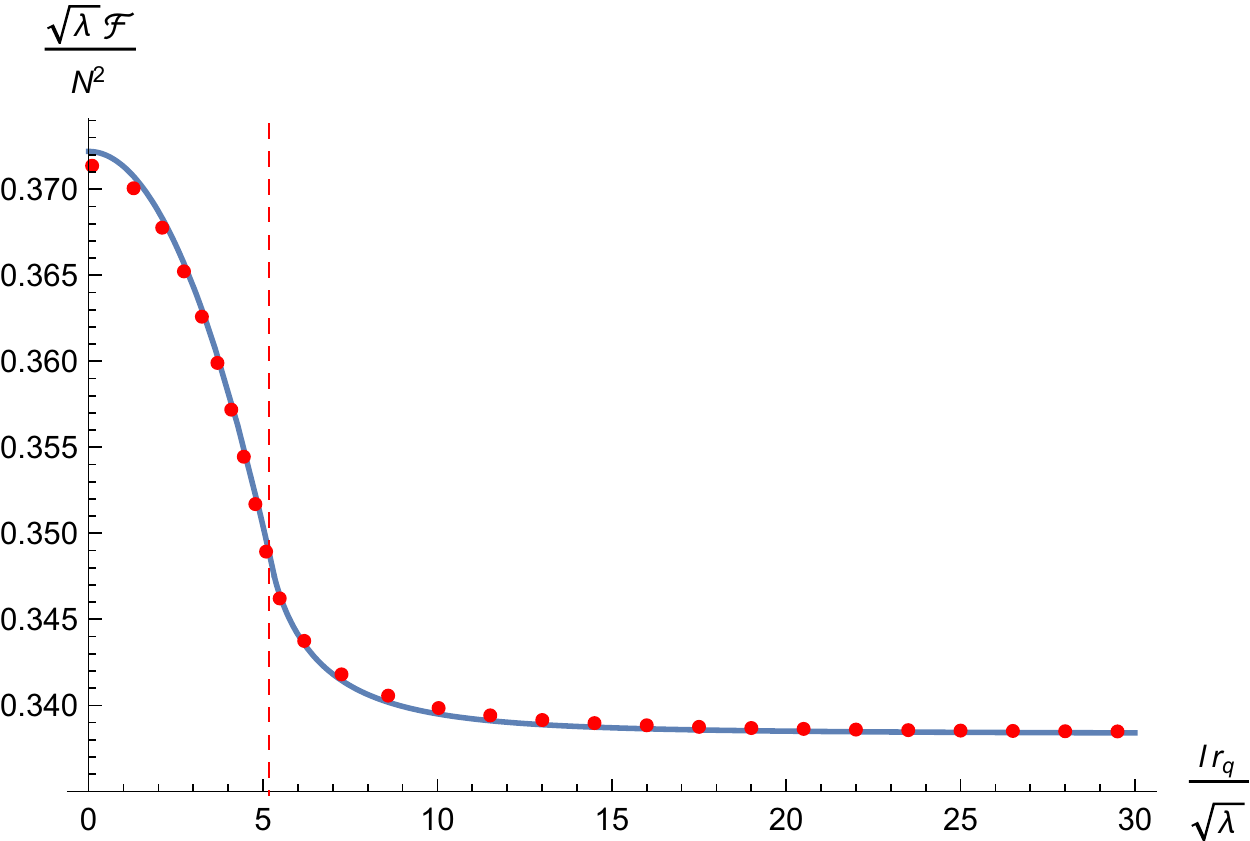}
   \caption{ $\mathcal{F}$  monotonically interpolates between constant values at the UV ($l r_q/\sqrt\lambda\to 0$) and IR ($l r_q/\sqrt\lambda\to\infty)$ conformal fixed points.  The vertical dashed line is  at $l_q r_q/\sqrt{\lambda}$ where $l_q$ is the width of the strip whose bulk minimal surface just touches the cavity. The points are from the full numerical solution and the continuous curve is from our analytic results in the few flavor limit (small $\hat\epsilon$).  Here $\hat{\epsilon}=0.1$.}
   \label{fig:LiuMezei}
\end{figure}

In quantum field theory one seeks to define a c-function that counts the number of degrees of freedom available as a function of  the scale of measurement $l$.  Liu and Mezei  \cite{Liu:2012eea, Liu:2013una} proposed that a c-function can be defined in terms of the derivative of the entanglement entropy with respect to scale.\footnote{A spatial integral of a similar quantity, called the differential entropy, was shown to reproduce the areas of closed surfaces in AdS space \cite{Balasubramanian:2013lsa,Headrick:2014eia}, see also \cite{Jones:2015twa}.}  In our context, where we are computing the entanglement of strips, we can write this as
\begin{align}
   \mathcal{F}(l) = {l^2 \over L_y} \frac{\partial S^{reg}}{\partial l}.
   \label{eq:liumezei}
\end{align}
where $l$ is the strip width and $L_y$ is the strip length which can be taken to infinity.\footnote{In more detail, following \cite{Liu:2013una}, the entanglement entropy of the strip will have the general behavior $S \sim L_y ( {\rm divergent} + {\rm finite})$.   The divergence comes from short distance modes straddling the edges of the strip and so will be proportional to $L_y$, but will not depend on the strip width $l$.  Thus we can extract the finite part by computing $l \partial S /\partial l$.  However, this quantity will diverge with the length of the strip so we should divide by $L_y$ to get a finite entanglement per unit length as $(l/L_y) \partial S /\partial l$.    We can make this quantity dimensionless by multiplying by $l$, to define $\mathcal{F} = (l^2/L_y)  \partial S /\partial l$.  At the conformal points there is no scale, so $\mathcal{F}$ should be a constant, independently of the strip width $l$, but away from conformality $\mathcal{F}$ will be a dimensionless combination of the scales in the theory.}   $\mathcal{F}(l)$ is UV finite and interpolates in a renormalizable QFT between fixed, scale-independent values at the UV and IR fixed points.   We can compute $\mathcal{F}(l)$ analytically using our explicit expressions for entanglement entropy as a function of strip width.   Plotting the results (Fig.~\ref{fig:LiuMezei}) indeed shows that $\mathcal{F}(l)$ in our massive ABJM theory approaches a constant value in the UV (small $l r_q/\sqrt\lambda$) and decreases monotonically to another constant in the IR (large $l r_q/\sqrt\lambda$), consistently with the interpretation that $\mathcal{F}(l)$ counts the degrees of freedom at a scale $l$. Similar statement also holds for disk regions in the same geometry \cite{Bea:2013jxa}.

In our model ${\mathcal F}$ is extremized at the fixed points of the flow, {\emph{i.e.}}, the first derivative $\partial {\mathcal F} / \partial l$ will vanish in the large and small $l$ limits.   
Since ${\mathcal F}$ is both continuous and monotonic, the first non-vanishing derivatives should be negative in the UV ($l r_q/\sqrt\lambda \to 0$) and positive in the IR ($l r_q/\sqrt\lambda\to\infty$).  
Thus, we expect the second derivative ${\mathcal{F}}''(l)$ to change sign at some intermediate scale.  In the few flavor limit (small $\hat\epsilon = (3/4)(N_f/N)\lambda$) we can show explicitly that
\begin{align}\label{eq:Fpp}
   \frac{\partial^2 \mathcal{F}(l)}{\partial l^2} = \begin{cases} - \hat{\epsilon} \frac{\Gamma\left(\frac{1}{4}\right)^4}{72\sqrt{2}\pi^4} \frac{N^2 r_q^2}{\lambda^{3/2}} & , \
l \to l_\text{crit}^-
   \\
      \hat{\epsilon} \frac{\Gamma\left(\frac{1}{4}\right)^2}{12\pi^{7/2}} \frac{N^2 r_q^2}{\lambda^{3/2}} \left( \frac{l}{l_\text{crit}} - 1 \right)^{-1/2} & ,  \
  l \to l_\text{crit}^+    \ .
   \end{cases}
\end{align}
Notice that since we are working at first order in $\hat\epsilon$, $l_\text{crit}$ in (\ref{eq:Fpp}) can be interpreted as $l_q$, {\emph{i.e.}}, as the width of the strip whose bulk minimal surface just touches the cavity in the gravity solution.    
In other words, ${\mathcal{F}}''(l)$  changes sign at a scale corresponding to the mass of the fundamental quarks.   Note also that the second derivative is discontinuous across the cavity. This is expected, as the Einstein equation of our background solution has a discontinuity in the source at the location of the cavity where the energy-momentum tensor is abruptly turned on. One can refine the smearing procedure \cite{Conde:2011rg} by using embedded D6-branes with  a distribution of  tip positions that are not delta-function peaked at $x_*$. This would lead to a continuous topological transition in the geometry at the scale $r_q$ with smooth second derivatives of the background metric and thus of ${\mathcal{F}}(l)$.

\begin{figure}[!ht]
   \centering
   \includegraphics[width=0.6\textwidth]{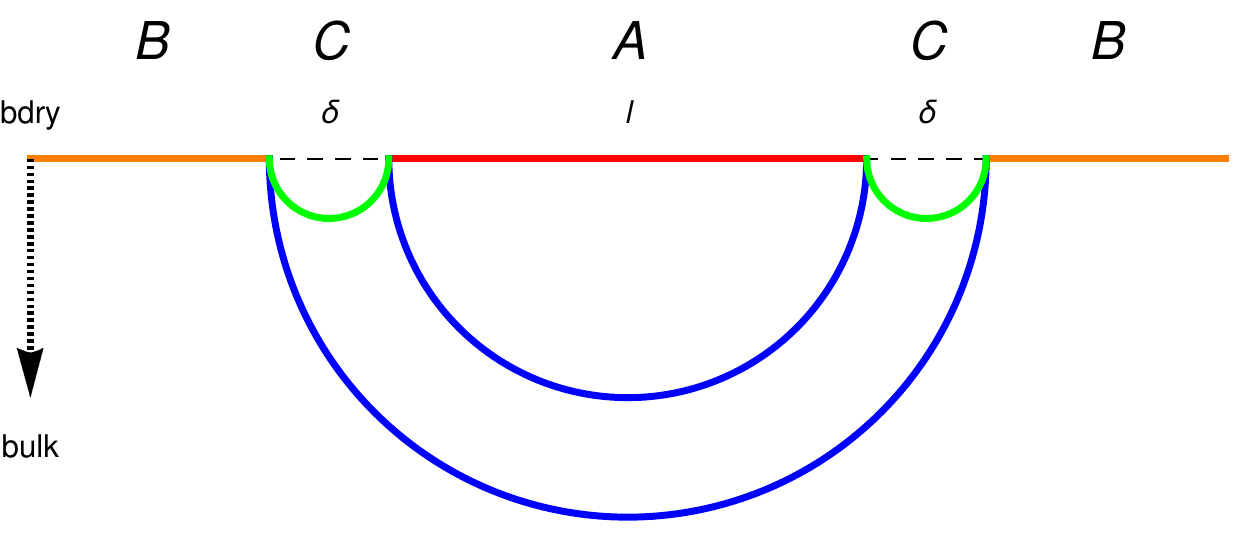}
   \caption{Strip configuration we consider in the definition of the c-function. We have a strip of width $l$ bordered by two thin strips of width $\delta$. We consider the mutual information between the center strip (red) and the exterior (orange).}
   \label{fig:mutual_info_setup}
\end{figure}

Casini et al.  \cite{Casini:2015woa} proposed an alternative method of computing a c-function for three-dimensional QFTs  from the  mutual information $I(A_+,A_-)$ between the interior of a smaller circle ($A_-$) and the exterior of a larger circle ($A_+$) with an annulus of width  $\delta$ between them. 
The advantage of working with mutual information is that this quantity is UV-finite, and is thus well-defined without any regularization unlike entanglement entropy.  Since we are working with strips, not circular regions, we consider the mutual information between a strip of width $l$ (region A) and the remainder of the space (region B) outside a boundary region of width $\delta$ on either side of the strip (region C) -- see  Fig.~\ref{fig:mutual_info_setup}. We will take $\delta \to 0$ in the end.
We then define our candidate c-function to be 
\begin{align}
   c(l) = \lim_{\delta\to 0} \frac{l^2}{L_y} \frac{\partial I(l,\delta)}{\partial l} \ , \label{eq:c-function}
\end{align}
where the mutual information is given by $I(A,B) = S(A)+S(B)-S(A\cup B)$, where $S(X)$ is the entanglement entropy of region $X$. 
In fact, we can show that 
\begin{align}
   c(l) = 2\mathcal{F}(l) \ .
\end{align}
To this end, consider a QFT  partitioned into three regions -- A, characterized by length scale $l$; C, bordering $A$ characterized by a size $\delta \ll l$; and B, which is the rest of the space.
 We now write our first candidate c-function (\ref{eq:liumezei}) in terms of the differential operator  $L_l = (l^2 / L_y) \partial/\partial l$ as
\begin{align}
   \mathcal{F}(l) = L_l S(A) \ .
\end{align}
Our second candidate c-function is 
\begin{align}
   c(l) = \lim_{\delta\to 0} L_l I(A,B)\ .
\end{align}
The equality $c=2\mathcal{F}$ then follows for pure states:
\begin{align}
   c(l) &= \lim_{\delta\to 0} L_l S(A) + \lim_{\delta\to 0} L_l S(B) - \lim_{\delta\to 0} L_l S(A\cup B) \\
   &= L_l S(A) + \underbrace{\lim_{\delta\to 0} L_l S(B)}_{= L_l S(A)} - \underbrace{\lim_{\delta\to 0} L_l S(C)}_{=0}= 2 \mathcal{F}(l) \ .
\end{align}
Here we used the fact that $C$ is the complement of $A\cup B$ and so $S(C) = S(A\cup B)$ for pure states.    As $\delta$ vanishes, so does the ``bulk'' contribution to $S(C)$, so that $S(C)$ becomes equal to its UV-divergent piece.  The latter is proportional to  the length of the boundary between $A$ and $B$ and does not depend on $l$.  Thus $ L_l S(C)$ vanishes in the $\delta \to 0$ limit. Likewise,  $L_l S(B) \to L_l S(A)$ as $\delta \to 0$, since $B$ is the complement of $A$ in this limit, and $S(A) = S(\bar{A})$ for pure states.  Note that, as discussed earlier, the differential operator $L_l$ removes the divergences in the strip entanglement entropy so that these arguments are well-defined.  We stated this proof for strips, but a similar argument holds for compact $A$.


\subsection{Flow of extensivity}\label{sec:extensivity}

Consider three entangling regions $A$, $B$, and $C$ and their tripartite information
\begin{equation}
I_3(A,B,C) = I(A,B) + I(A,C) - I(A, B \cup C) \ .
\end{equation}
Despite appearances, this quantity is symmetric between $A$, $B$, and $C$ as can be verified by expressing the mutual informations in terms of entanglement entropies.  $I_3$ is a measure of extensivity of mutual information in the sense that $I_3 = 0$ implies that the information that $A$ shares with $B \cup C$ is the sum of the information shared with $B$ and $C$ separately. In a general quantum field theory $I_3$ can take either sign depending on the choice of state and entangling regions \cite{Casini:2008wt}, but it was shown in  \cite{Hayden:2011ag} that in holographic theories $I_3 \leq 0$.   This means that the information that region $A$ shares with $B$ and $C$ is in general super-extensive in a holographic theory -- there is information in intrinsically 3-party entanglement that cannot be uncovered just from the 2-party entanglement.  To investigate how the extensivity of mutual information varies with scale we define the ratio
\begin{align}
   e = \frac{I(A,B \cup C)}{I(A,B)+I(A,C)} \ , \label{eq:ext_definition}
\end{align}
so that $e = 1$ in an extensive theory, $e>1$ in a super-extensive theory, and $e<1$ in a sub-extensive theory.    The extensivity $e$ is only well defined when the denominator is non-vanishing, {\emph{i.e.}}, if the region $A$ shares at least some mutual information with  at least one of $B$ or $C$ separately.\footnote{Note that the mutual information is a non-negative quantity so that the two terms in the denominator cannot cancel each other.}   As we have seen, in some entangling region configurations the mutual information can be identically zero.  If that happens for $I(A,B)$ and $I(A,C)$ while $I(A,B \cup C)$ remains positive then $e$ diverges, representing maximal super-extensivity, and this can certainly happen in quantum theories.\footnote{See an interesting example in the context of multi-boundary wormholes in \cite{Balasubramanian:2014hda}.}   Below we will first determine configurations where $e$ is finite for a CFT, and then ask how the massive ABJM theory behaves in these situations.

\begin{figure}[!t]
   \centering
   \includegraphics[width=.8\textwidth]{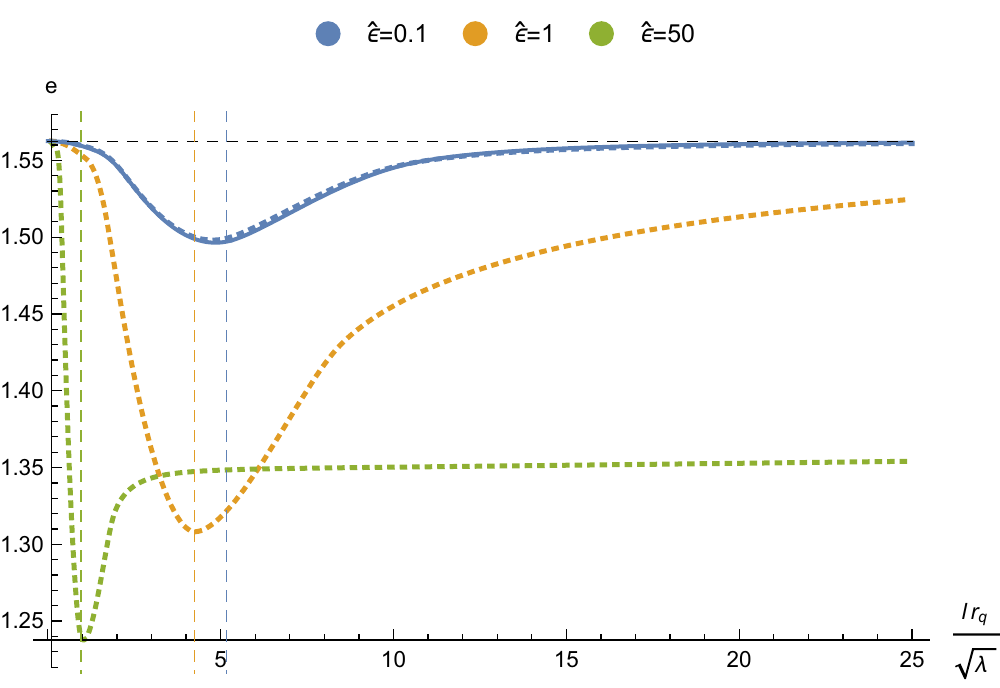}
   \caption{
   Extensivity ($e$) as a function of the width $l$ of the entangling strips shown for a fixed ratio of strip separation to strip width ($s/l=0.5$). The extensivity only depends on the dimensionless combination $l r_q/\sqrt{\lambda}$. The black, dashed horizontal line is the result in a conformal background ($\hat\epsilon = 0$ or massless quarks) and $s/l = 0.5$, and is thus the limit approached in the deep UV ($ l \to 0$) and deep IR $l \to \infty$ for any quark mass. Vertical dashed lines are $l_qr_q / \sqrt{\lambda}$ where $l_q$ is the width of the strip whose bulk minimal surface just touches the cavity for each $\hat\epsilon = (3/4) (N_f/N)\lambda$, and thus represent the scale where quarks have the largest effect on the dynamics. Dashed lines are numerical results and the solid line is our analytical small $\hat{\epsilon}$-expansion.   
    }
   \label{fig:ext_grid_new}
\end{figure}

Consider three strips $A$, $B$, and $C$ of equal width $l$ that are separated by distance $s$ with $A$ in the middle.  We can study how the extensivity of entanglement changes with scale by varying $l$ while keeping $s/l$ fixed (Fig.~\ref{fig:ext_grid_new}).   In the absence of massive quarks the theory is conformal, and so the extensivity $e$ is constant under this scaling variation.  Applying the standard formulas for the entanglement entropy of intervals in a $(2+1)$-dimensional CFT (see, {\emph{e.g.}}, \cite{Ryu} or the first term on the right hand side of Eq.~(\ref{eq:SUV})) one finds a simple result:
\be\label{eq:eCFT}
 e(s/l) = \frac{1}{2}\cdot\frac{3 (s/l)^3+8 (s/l)^2+(s/l)-6}{2 (s/l)^3+5 (s/l)^2+(s/l)-3}  \ , \ {\text{for a CFT}} \ .
\ee
This formula applies if the denominator is positive, which is true so long as  $ s/l < (\sqrt 5-1)/2 = 1/\varphi$.  The extensivity $e$ increases monotonically from $e=1$ at $s/l = 0$ to $e=\infty$ at $ s/l = 1/\varphi$.  At the latter point  the strip $A$ only shares information with $B$ and $C$ in an intrinsically tripartite manner, so that the theory is maximally super-extensive.  We are therefore going to compare the CFT with the massive ABJM theory in the range $s/l <  1/\varphi $.

We see from Fig.~\ref{fig:ext_grid_new} that when massive quarks are present the mutual information comes closest to adding extensively ($e = 1$) at an intermediate scale, while approaching the conformal value of $e$ in the UV and IR.  Thus massive quarks seem to make entanglement structure of the vacuum  more bipartite.
As $\hat\epsilon \to \infty$, the super-extensive behavior declines rapidly away from the UV conformal limit, reaches a minimum, and then increases steeply again to an intermediate value that remains essentially constant for a large range of scales.  The right hand panel in Fig.~\ref{fig:peak_height} shows a similar saturation of the critical scale for mutual information transitions at large $\hat\epsilon$. This  suggests that the many flavor limit  ($N_f \to \infty$) may manifest a phase transition to a new regime where the theory is not conformal but nevertheless has (nearly) constant extensivity at all scales.

\begin{figure}[!t]
   \centering
   \includegraphics[width=0.49\textwidth,height = 2.2in]{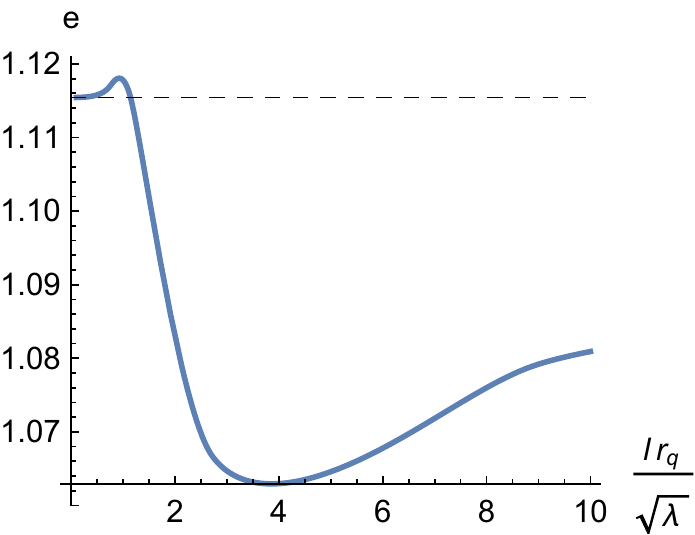} \hfil   
      \includegraphics[width=0.49\textwidth,height = 2in]{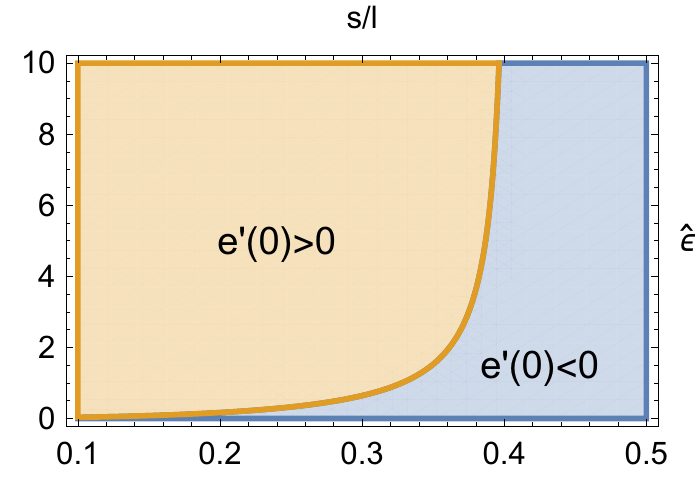} \\
      ({\bf A}) \hfil   \hspace{0.75in}  ({\bf B})  \hfil 
   \caption{
   ({\bf A}) Extensivity $e$ as a function of strip width $l$, with a fixed ratio of strip separation to strip width ($s/l=0.3$); here $e$ is calculated numerically for $\hat{\epsilon}=(3/4)(N_f/N)\lambda = 5$ and the horizontal dashed black line corresponds to the analytical result (\ref{eq:eCFT}). Note the small range of scales where the theory is {\it more} super-extensive than in the conformal UV ($l \to 0$) and IR ($l \to \infty$) limits.
  ({\bf B}) Domains in the $\{\hat\epsilon,s/l\}$ parameter plane where the theory display greater super-extensivity than a CFT ($e'(0)> 0$) or is always less super-extensive than a CFT ($e'(0)<0$). Analytical results indicate that for $\hat\epsilon=0$, $e'(0) <0$ for all $s/l$ and in the $\hat\epsilon\to\infty$ limit the derivative changes sign at $s/l\sim 0.406$ (the solution of (\ref{eq:ext_derivative_UV}) for $b=5/4$).   
  }
     \label{fig:ext_nonmonotonous}
\end{figure}

These results suggest that conformal field theories might be maximally super-extensive.  Indeed, we might suspect that the absence of a scale in a conformal field theory would make it easier for  intrinsically multi-party quantum correlations to develop between widely separated regions, making mutual information as super-extensive as possible.   Thus, it is a surprise to observe that in our RG flows there is a narrow range of scales near the UV limit in which the ABJM theory with massive quarks is {\it more} super-extensive than the conformal theory (Fig.~\ref{fig:ext_nonmonotonous}A).      To test for this we can simply ask whether $\partial_l e(l) > 0 $ in the conformal, double-scaling limit where $l \to 0$ with $s/l$ fixed.   Since this is deep in the UV region, we can analyze it holographically in an asymptotic expansion in inverse powers of the radial coordinate in the dual geometry, getting a result that is accurate to all orders in $\hat\epsilon$ (details in Appendix~\ref{app:UVexpansion}).  
A straightforward calculation setting $\partial_l e |_{l=0}  = 0$  gives the condition
 \begin{gather}
    -\left(\frac{s}{l}\right)^{2 b}-2 \left(\frac{s}{l}\right)^{2 b+1}-\left(\frac{s}{l}\right)^{2 b+2}+3 \left(\frac{s}{l}\right)^2 \left(\left(\frac{s}{l}\right)+2\right)^{2 b} \nonumber \\ -\left(\left(\frac{s}{l}\right)^2+\left(\frac{s}{l}\right)-1\right) \left(2 \left(\frac{s}{l}\right)+3\right)^{2 b}+2 \left(\frac{s}{l}\right) \left(\left(\frac{s}{l}\right)+2\right)^{2 b}-3 \left(\left(\frac{s}{l}\right)+2\right)^{2 b} \nonumber \\ -\left(\frac{s}{l}\right)^2+\left(\frac{s}{l}\right)+3 = 0 \ . \label{eq:ext_derivative_UV}
\end{gather}
Here $b$ is a function of $\hat\epsilon = (3/4) (N_f/N) \lambda$ through the definition (\ref{b_new}).  Solving this equation leads to Fig.~\ref{fig:ext_nonmonotonous}B, which reveals two domains in the $\{ \hat\epsilon, s/l \}$ plane, one in which the extensivity $e$ increases from the conformal point, and one in which it does not.
We see that for any number of flavors ({\emph{i.e.}}, any $\hat\epsilon > 0$), $e'(0) = \partial_l e |_{l=0}$ is positive for small enough $s/l$ so that the theory must be more super-extensive than in the conformal limit for some range of scales near the UV limit.


\subsection{Flow of conditional mutual information}
\label{sec:conditional}

We can also characterize the sharing of quantum information across space in terms of conditional entanglement measures that quantify the amount of uncertainty about one spatial region conditioned on knowledge of others.  For example, one can define the conditional entanglement entropy between regions $A$ and $B$ as
\begin{align}
   H(A|B) = S(A \cup B) - S(B)\ .
\end{align}
Famously, one can show that $S(A) \geq H(A|B)$ implying that conditioning cannot increase entropy, even if the conditioning system is quantum mechanical.   The conditional entropy can be negative, which implies that in quantum theories one can sometimes be more certain about the joint state of the system than about the states of the constituent parts.   States with negative conditional entropy are necessarily non-separable.    Thus, the flow of conditional entropy with the size of the entangling systems should probe how quantum information is organized at different spatial scales.  However, in quantum field theory $H(A|B)$ is UV divergent.  Thus, to avoid the subtlety of regularizing $H(A|B)$ we turn instead to the conditional mutual information.

Mutual information between $A$ and $B$ conditioned on $C$ is defined as
\begin{align}\label{eq:conditioned}
   I(A,B|C) = S(A\cup C) + S(B \cup C) - S(C) - S(A \cup B \cup C) \ ,
\end{align}
and is always non-negative due to strong subadditivity in holographic settings\cite{Headrick:2007km,Bao:2015bfa,Freedman:2016zud}.  
Conditional mutual information is related to the extensivity parameter introduced in the previous section as
\begin{align}
   e = \frac{I(A,B|C)+I(A,C)}{I(A,B)+I(A,C)} \label{eq:extensivity_conditioned} \ .
\end{align}
We will consider the conditional mutual information $I(A,B|C)$ for three strips of equal width $l$, that are separated by intervals of width $s$ and ordered from left to right as $A$, $B$, and $C$.  In other words, we will ask how much information is shared between neighboring strips $A$ and $B$ conditioned on knowledge of a third strip $C$ that is further to the right.\footnote{We checked that qualitatively similarly results were obtained when the strip $C$ was positioned in between $A$ and $B$.}

\begin{figure}
   \centering
   \includegraphics[width=\textwidth]{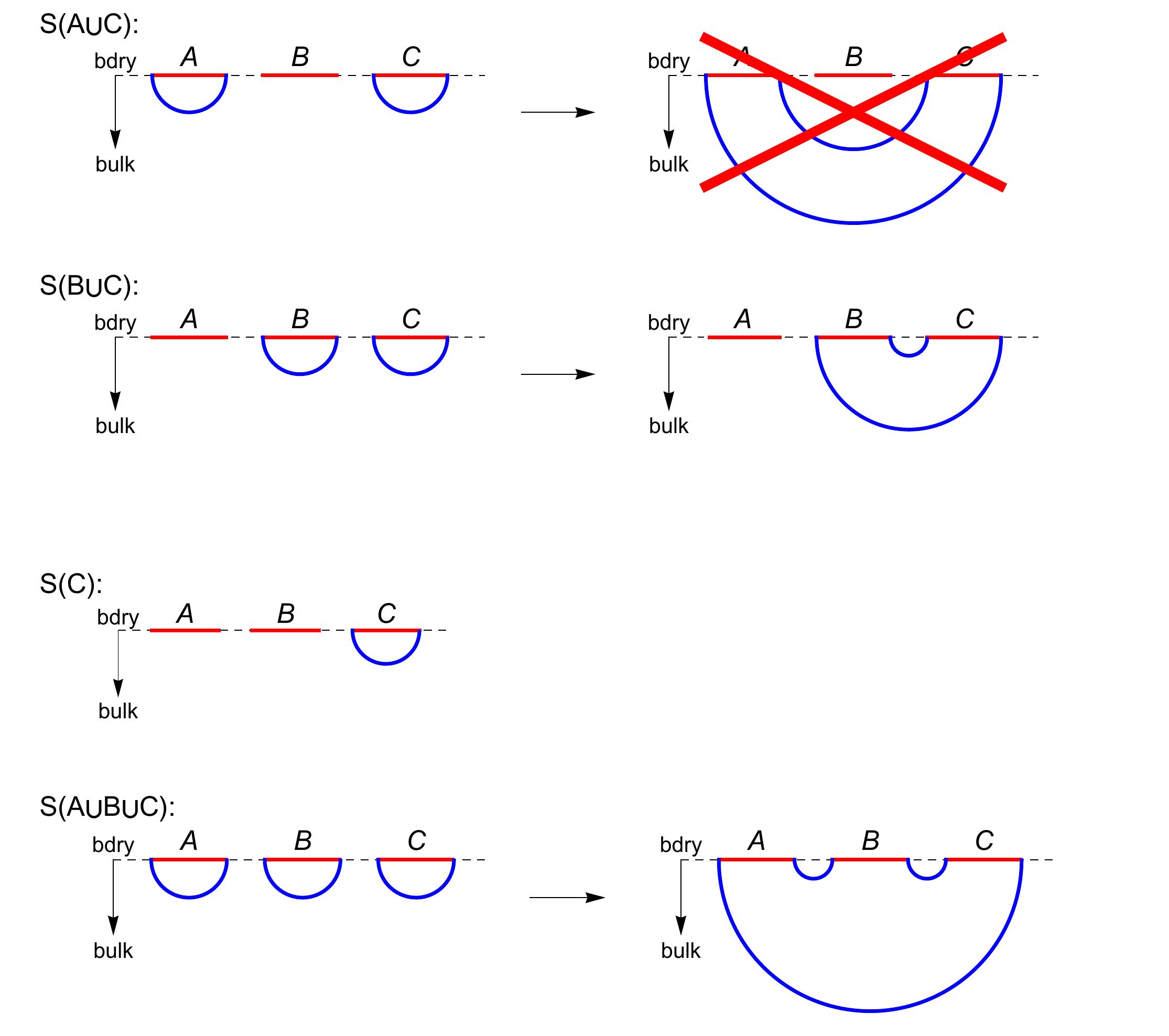}
   \caption{Schematic of transitions in the entropies contributing to the conditional mutual information. Here, $A$, $B$, and $C$ are strips of width $l$ separated by gaps of width $s$.  The blue arcs in figure shows the minimal surfaces in the holographic geometry that compute  the entanglement entropy for the union of the indicated regions.  The ``disconnected'' phases on the left occur when $s/l$  is large enough.  In these phases the the entanglement entropy of a union of regions equals the sum of the entanglement entropies of the regions separately.   When $s/l$ is small enough there is a transition to the ``connected'' phases indicated on the right.  In the connected phases the entanglement entropy of a union of regions is {\it not} equal to the sum of the entanglement entropies of the regions separately because they share mutual information.   The connected phase for $S(A \cup C)$ is crossed out since  that phase never occurs in our case.  This is because $A$ and $C$ are always separated by an interval of length at least $l$.}
   \label{fig:Icond_phase_graph}
\end{figure}

Holographically, the conditional mutual information in (\ref{eq:conditioned}) is computed in terms of minimal bulk AdS surfaces that span the CFT domains $C$, $A \cup C$, $B \cup C$, and $A \cup B \cup C$.   Depending on the relative size of $l$ and $s$ the minimal surfaces for the latter three domains either extend between the boundaries of the disjoint regions $A$, $B$, $C$ (connected phase), or localize to span $A$, $B$, and $C$ separately (disconnected phase).  The phases are illustrated in Fig.~\ref{fig:Icond_phase_graph}.  (This happens because of the mutual information transitions discussed in Sec.~\ref{sec:mutinfotrans}.)   The conditional mutual information therefore behaves differently on the different sides of these transitions.

\begin{figure}[!ht]
   \centering
   \includegraphics[width=0.48\textwidth]{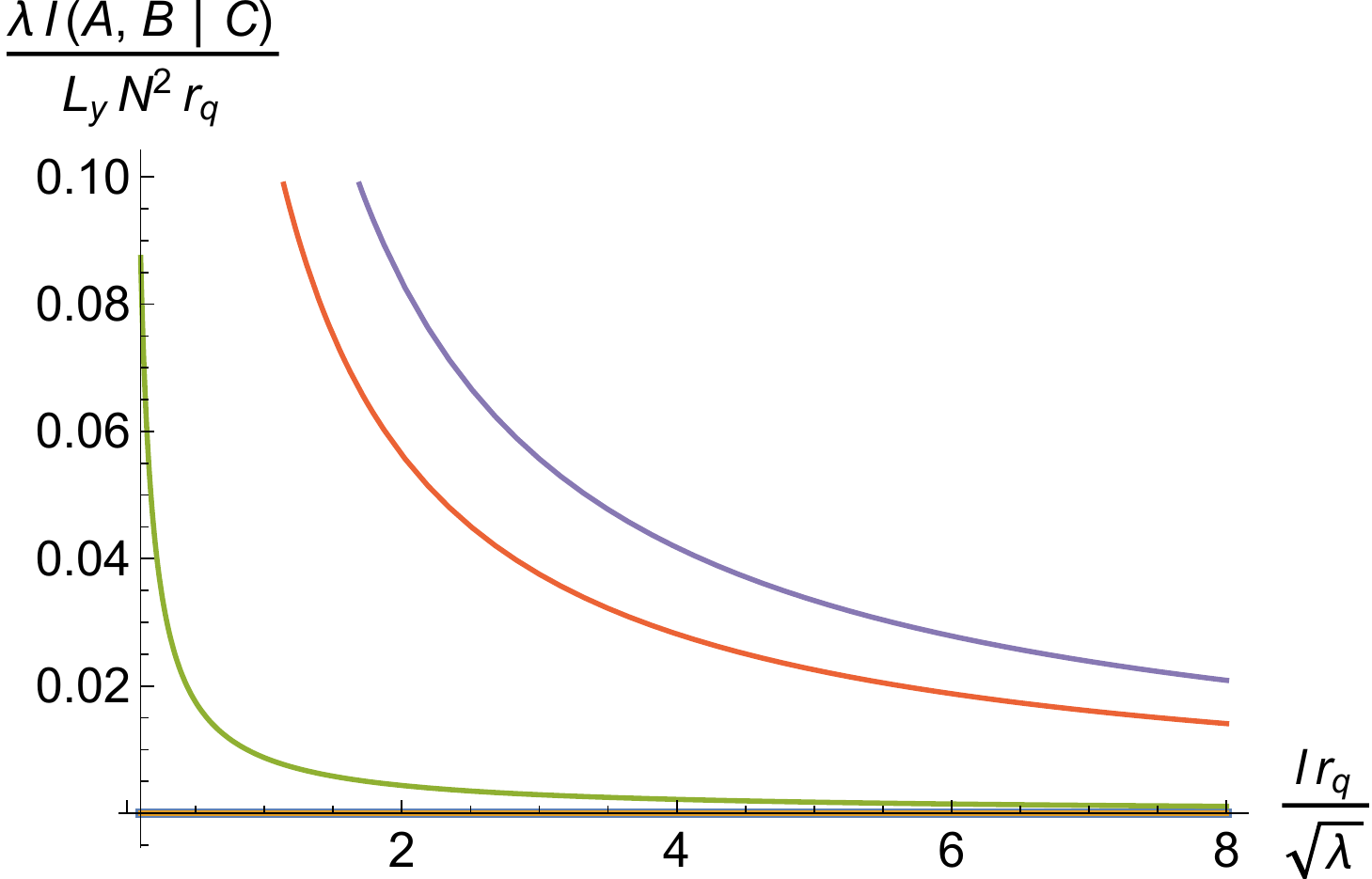}\includegraphics[width=0.48\textwidth]{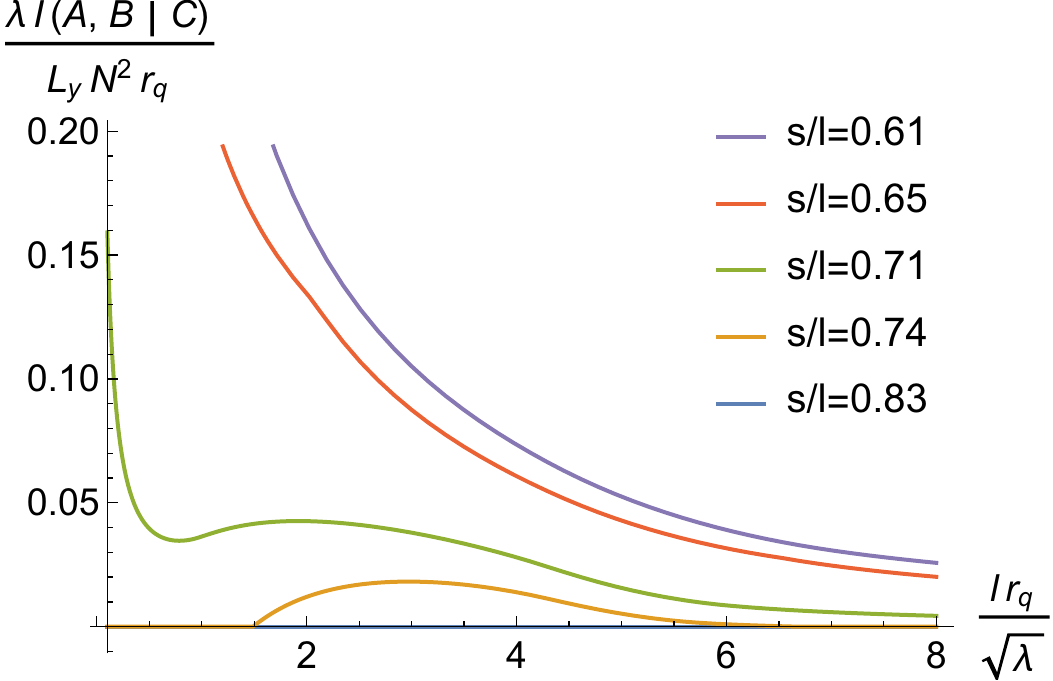} \\
    ({\bf A}) \hfil   \hspace{0.75in}  ({\bf B})  \hfil 
   \caption{Conditional mutual information $I(A,B|C)$ for strips $A$, $B$, and $C$ separated by spacing $s$, as a function of changing strip width $l$ with $s/l$ fixed.  The strips are ordered with $B$ in the middle as shown in Fig.~\ref{fig:Icond_phase_graph}.
({\bf A}) ABJM theory without quarks.  ({\bf B}) ABJM theory with massive quarks and $\hat\epsilon = (3/4) (N_f/N)\lambda = 1$.
   }    \label{fig:conditioned}
\end{figure}

First consider the theory without quarks that is holographically described as empty AdS$_4$.  There is no intrinsic scale in this theory, so the holographic computation of mutual information displays  two possible configurations: either the strips are sufficiently far apart, the bulk minimal surfaces are disconnected, and $I(A,B|C)=0$; or the strips are close, the bulk minimal surfaces are connected, and $I(A,B|C)$ monotonically decreases as a function of the strip width. This leads to results in Fig.~\ref{fig:conditioned}A.  At sufficiently large $s/l$ the conditional mutual information vanishes because all entanglement entropies in (\ref{eq:conditioned}) are in the holographically disconnected phase (orange and blue lines coinciding with the horizontal axis).   As $s/l$ decreases, the entropy $S(A \cup B \cup C)$ transitions to a phase described holographically by a connected bulk minimal surface (Fig.~\ref{fig:Icond_phase_graph}, bottom row) while the surfaces for $S(A \cup C)$ and $S(B \cup C)$ remain disconnected.   In this phase the conditional mutual information decreases monotonically with $l$  at fixed $s/l$ (green line in Fig.~\ref{fig:conditioned}A).  Finally, at sufficiently small $s/l$, the minimal surface for $S(B\cup C)$ also becomes connected (Fig.~\ref{fig:Icond_phase_graph}, second row).   Again in this phase, the conditional mutual information decreases monotonically with the width of the strips (red line in Fig.~\ref{fig:conditioned}A). There is never a phase where the minimal surface for $S(A\cup C)$ becomes connected because $A$ and $C$ are always separated by a distance greater than the strip width $l$.

Now consider the ABJM theory with massive quarks (Fig.~\ref{fig:conditioned}B).   At sufficiently large separations of the strips relative to their width (large $s/l$), the conditional mutual information vanishes like in a CFT (blue line in Fig.~\ref{fig:conditioned}B), since all its constituent entanglement entropies are in their disconnected phases.  
Likewise, for sufficiently small $s/l$, the conditional mutual information declines monotonically from the UV (small $l$) to the IR (large $l$), like in a CFT (red and purple lines in Fig.~\ref{fig:conditioned}B).   But for intermediate $s/l$ deviations from CFT-like behavior are pronounced.  In particular, as $s/l$ decreases from large values there is a critical value at which $S(A \cup B \cup C)$ enters the connected phase in an intermediate range of  $l$, $l_0 \leq l \leq l_1$.\footnote{One can see that in the massive ABJM theory the connected phase for mutual information typically occurs in a restricted range of strip widths $l$ for any fixed ratio $s/l$ by examining Fig.~\ref{fig:peak_height}A.  In this figure a horizontal line of fixed $s/l$ will cut through the critical curve at two values of $l$; the connected phases lies in between these two points.}
 This leads to conditional mutual information curves like the orange line in Fig.~\ref{fig:conditioned}B ($s/l = 0.74$), where, $I(A,B|C)=0$ for small $l$, but increases sharply from 0 at a critical value of $l$, reaches a maximum, and then decays to 0 again at large $l$.  This non-monotonic behavior stands in sharp contrast to the behavior in a CFT, where, for a given $s/l$, $S(A \cup B \cup C)$  is either in the connected phase or the disconnected phase independently of the actual value of $l$.   In general the non-monotonicity of conditional mutual information for a theory with an intrinsic scale is not only driven by phase transitions in the mutual information.  For example, in the green curve in Fig.~\ref{fig:conditioned}B ($s/l = 0.71$), $S(A \cup B \cup C)$ is in the connected phase for all $l$ and $S(B\cup C)$ is always disconnected, but $I(A,B|C)$ is nevertheless non-monotonic because of the functional dependence of the entanglement entropy itself on $l$ ({\emph{i.e.}}, it does not scale simply as $1/l$ except in the deep UV and IR).  Meanwhile, in the red curve ($s/l = 0.65$) $S(A \cup B \cup C)$ is always in the connected phase while  $S(B\cup C)$ is connected only for an intermediate range of  $l$, but despite these transitions in $S(B\cup C)$ the  conditional mutual information is monotonically declining.  Finally, for sufficiently small $s/l$ (purple curve, $s/l = 0.61$) , both  $S(A \cup B \cup C)$  and $S(B\cup C)$ are always in the connected phase and the conditional mutual information decreases monotonically with $l$.


\section{Discussion} \label{sec:conclusions}

In this paper we used holography to analytically explore the spatial distribution of quantum information in a strongly interacting Chern-Simons theory with massive quarks that  flows between conformal fixed points in the UV and the IR.   We computed entanglement entropy of strips in this theory by applying the Ryu-Takayanagi formula to the analytically known ten-dimensional gravitational dual spacetime.  From this entanglement entropy we constructed a c-function that flows monotonically from the UV to the IR.

To test how information is shared across space at different scales we computed the mutual information between strips of width $l$ separated by spacing $s$.   In holographic theories such configurations manifest sharp transitions as a function of $s$ and $l$ on one side of which the mutual information vanishes.  We found that for a range of scales $l$, this transition occurs at a separation $s$ between strips that is greater than in a CFT.  This critical separation is maximized at a scale related to the inverse quark mass.   Interestingly, this suggests that information is more non-locally shared in some regimes away from the conformal fixed points.  This result also implies that in a certain range of $s/l$ the mutual information between strips vanishes for both small and large $l$ but is non-zero at intermediate scales.  This behavior differs markedly from a CFT where the mutual information changes monotonically with scale, and suggests that, in our theory, two strips with a fixed separation that share mutual information can sometimes be mutually disentangled by  increasing the strip width.  It is surprising that adding degrees of freedom to two entangled regions would seem to reduce their shared information in this way.   We also compared extensivity of mutual information in our theory with the super-extensive behavior in a CFT.  At most length scales the theory with massive quarks is less super-extensive than a CFT.  However, there is a range of $s$ and $l$ near the UV limit in which the massive ABJM is more super-extensive than a CFT.  This is surprising because the scale-free nature of CFTs might have led us to suspect that CFTs would be maximally super-extensive.

As a consistency condition on our results we can check the holographic entropy cone inequalities of \cite{Bao:2015bfa}.   For 2, 3, and 4 regions, strong subadditivity and monogamy of mutual information give the complete set of inequalities for CFT states with smooth holographic dual geometries. For 5 or more regions, there are new inequalities. One of these is a family of cyclic inequalities
\begin{align}
   \sum_{i=1}^n S(A_i | A_{i+1} \dots A_{i+k}) \geq S(A_1 \dots A_n) \ ,
\end{align}
where $n$ is the number of regions and $n=2k+1$. We have numerically checked this family of inequalities in massive ABJM theory for parallel strip configurations up to 9 separate regions.

The results in this paper depended on having a fully ten-dimensional gravitational dual to a strongly coupled QFT.  In fact, the extremal surfaces computing entanglement entropy explored the full 10 dimensions.  Many recent applications of holography to the study of quantum information have treated the gravitational dual as effectively having one extra dimension compared to the field theory.   But a fully realized duality in string theory requires 10 dimensions, and, as we have seen here, the details of the full geometry manifest themselves in the structure of the RG flow.   Indeed, as discussed in \cite{Jones:2016iwx}   the ``internal'' and AdS dimensions can interact significantly, and even sometimes completely exchange roles \cite{Balasubramanian:2017hgy}.   It would be useful to develop more such examples. 
Finally, it would be interesting to put the Chern-Simons field theory studied in this paper on a 3-sphere and test the recent proposal for the c-function \cite{Ghosh:2018qtg}.


\paragraph{Acknowledgments}

\noindent
We would like to thank Matt DeCross, Arjun Kar, Esko Keski-Vakkuri, Onkar Parrikar, and G.~S\'{a}rosi for helpful discussions. A.~V.~R. is funded by the Spanish grants FPA2014-52218-P and FPA2017-84436-P by Xunta de Galicia (GRC2013-024), by FEDER and by the Maria de Maeztu Unit of Excellence MDM-2016-0692.   V.~B. was supported in part by the Simons Foundation (\# 385592, V.~B.) through the It From Qubit Simons Collaboration, and the US Department of Energy grant FG02-05ER-41367. V.~B. also acknowledges the hospitality of the Aspen Center for Physics which is supported by National Science Foundation grant PHY-1607611.    This work was initiated during the ``Holographic methods for strongly coupled systems" workshop  of the Galileo Galilei Institute in Florence.


\appendix

\section{Details of the background}\label{app:details}

Here we will give the detailed expressions for the different functions appearing in our background (for a derivation see \cite{Bea:2013jxa}). Inside the cavity, {\emph{i.e.}, for $x\le 1$, these functions are analytic and depend on the constant $\gamma$ characterizing the running of the master function $W(x)$ in the sourceless region. The functions $f$, $g$, and the dilaton $\phi$ are given by:
\bear
{e^{f}\over r_q} & = & \frac{1+\sqrt{1+4\gamma}}{\sqrt 2}\frac{x}{\left[\sqrt{1+4\gamma x}(1+\sqrt{1+4\gamma x})\right]^{1/2}} \nonumber \\
{e^{g}\over r_q} & = &  \frac{1+\sqrt{1+4\gamma}}{2}\frac{x}{\sqrt{1+4\gamma x}} \nonumber \\
 k e^\phi  & = &  \left(1+\sqrt{1+4\gamma}\right)\frac{x}{1+4\gamma x}(r_q^4 h)^{1/4}  \ ,
\label{f_g_phi_inside}
\eear
where $r_q$ is the $r$ coordinate of the tip of the brane (see (\ref{eq:rq})). 
For $x\ge 1$ the master equation (\ref{eq:masterequation})  for $W(x)$ must be integrated numerically (although an analytic solution in powers of $\hat\epsilon$ can be obtained, see below). In terms of $W$ and  $\eta$, the different functions are:
\bea
{e^{f}\over r_q} & = &  \left(\frac{(1+\sqrt{1+4\gamma})^2}{2\sqrt{1+4\gamma}}\right)^{1/3}\sqrt{\frac{3x}{W'+4\eta}}W^{1/6}\exp\left\{\frac{2}{3}\int_1^x \frac{\eta(\xi)d\xi}{W(\xi)}\right\} \nonumber \\
{e^{g}\over r_q} & = & \left(\frac{(1+\sqrt{1+4\gamma})^2}{2\sqrt{1+4\gamma}}\right)^{1/3}\frac{x}{W^{1/3}}\exp\left\{\frac{2}{3}\int_1^x \frac{\eta(\xi)d\xi}{W(\xi)}\right\} \nonumber \\
 k e^\phi & = &   \left(\frac{(1+\sqrt{1+4\gamma})^2}{2\sqrt{1+4\gamma}}\right)^{1/3}\frac{12x (r_q^4 h)^{1/4}}{W^{1/3}(W'+4\eta)}\exp\left\{\frac{2}{3}\int_1^x \frac{\eta(\xi)d\xi}{W(\xi)}\right\} \ .
\label{f_g_phi_outside}
\eea
The constant $\gamma$ appearing in (\ref{f_g_phi_inside}) and (\ref{f_g_phi_outside}) is  obtained by the shooting technique described in the main text. The dilaton in these two equations has been written in terms of the warp factor $h$. For $x\le 1$ the function $h(x)$ can be found analytically:
\bear
 \frac{k}{N}\,r_q^4\, h  & = &  \frac{\pi^2}{2}(\sqrt{1+4\gamma}-1)^4\left(1+\frac{1}{4\gamma x}\right)\Bigg[\alpha + 24\log\left[\frac{\sqrt{4\gamma x}}{\sqrt{1+4\gamma x}+1}\right] \nonumber \\ 
  & & + \left(\frac{1}{2}+6\gamma x+\frac{1+(1-6\gamma x)\sqrt{1+4\gamma x}}{4\gamma x}\right)\frac{\sqrt{1+4\gamma x}+1}{\gamma^2 x^2}     \Bigg] \ ,
\eear
whereas for $x\ge 1$ it can be written in terms of $W(x)$ as:
\be
\frac{k}{N}\,r_q^4\, h  =  4\pi^2 r_q^4 e^{-g}\left(W'+4\eta\right)\left[\int_x^\infty \frac{\xi e^{-3g(\xi)}d\xi}{W(\xi)^2}\right] \ . \label{eq:h_outside}
\ee
The integration constant $\alpha$ that appears in the warping $h$ is  fixed by continuity at $x=1$: $\lim_{x\to 1^-}h(x)=\lim_{x\to 1^+}h(x)$. However, the expression for $\alpha$ is lengthy, so we do not write it here explicitly.

\subsection{Expansion in flavor}
\label{sec:flavorexp}
The master equation, and the corresponding functions of the ansatz, can be solved analytically in a power series expansion in the flavor deformation parameter $\hat\epsilon$\cite{Bea:thesis}. If we write:
\begin{equation}
 W(x)=\sum_{n=0}W_n(x)\hat\epsilon^n=2x+  W_1(x) \hat{\epsilon}+  W_2(x)\hat{\epsilon}^2+  W_3(x)\hat{\epsilon}^3+ {\cal{O}}(\hat\epsilon^4)\,\,,
 \label{eq:Wseries}
\end{equation}
then we get for the first three functions in (\ref{eq:Wseries}) in the region $x\le 1$:
\bear
 W_1(x) & = & {12\over 5}\,x^2 \nonumber \\
 W_2(x) & = & {8\over 875}\,(171\,-\,70\,x)x^2 \nonumber \\
 W_3(x) & = & {16\over 197071875}\,
 \big[235688\,+\,135135\,\big(35\,x\,-\,76)\,x\big] \ .
 \eear
 Moreover,  for $x\ge 1$ we have:
\bear
  W_1(x) & = & \frac{7 x^2 - 2}{2 x} - \frac{1}{10 x^3} \nonumber \\
  W_2(x) & = & \frac{13125 x^8+3500 x^6-3962 x^4-3360 x^4 \log x+300 x^2-35}{14000 x^7} \nonumber \\
  W_3(x) & = & -\frac{23}{208000 x^{11}}+\frac{89}{184800 x^9}-\frac{302}{39375 x^7}+\frac{8293}{98000 x^5}-\frac{2120717}{8400000 x^3}+\frac{37}{96 x} \nonumber\\
  & &  -\frac{\log (x) \left(10703 x^4+20160 x^4 \log (x)-3600 x^2+840\right)}{70000 x^7}-\frac{15 x}{32} \ .
 \eear
The constant $\gamma$ obtained from the shooting method can also be expanded in powers of $\hat\epsilon$,
with the result:
\beq
\gamma  =  \frac{2\, \hat{\epsilon} }{5}+\frac{228\, \hat{\epsilon}^2}{875}+
\frac{18855104 \,\hat\epsilon ^3}{591215625}+
{\cal{O}}(\hat\epsilon^4)~.
\label{gamma_expanded}
\eeq
The remaining functions of the ansatz can be obtained by plugging these expansions into (\ref{f_g_phi_outside}) and (\ref{eq:h_outside}). The expressions found in this way are lengthy and will not be reproduced here.


\section{UV expansion of the entanglement entropy}\label{app:UVexpansion}
In this section we work out the UV expansion of the strip entanglement entropy. The holographic entanglement entropy on a strip can be computed from equation (\ref{S_strip_total}). The UV expansion is given by
\begin{align}
   S(l) = S_\infty(l) + \delta S(l)\ ,
\end{align}
where $S_\infty(l)$ is the entanglement entropy computed at the UV fixed point and $\delta S(l)$ is the first perturbation around the UV. Explicitly, these are given by
\begin{align}
   S_\infty(l) &= -\frac{4\pi^2 F_{UV}(\mathbb{S}^3)}{\left[ \Gamma \left( \frac{1}{4} \right) \right]^4} \frac{1}{l} \\
   \delta S &= \frac{V_6 L_y}{2 G_{N}^{(10)}} \int_{x_*}^\infty \left( \left.\frac{\partial L}{\partial H}\right|_{UV} \delta H + \left.\frac{\partial L}{\partial H_*}\right|_{UV} \delta H_* + \left.\frac{\partial L}{\partial G}\right|_{UV} \delta G \right) dx\ .
\end{align}
The UV expansions of $H(x)$ and $G(x)$ are

\begin{align}
   H(x) &= H_0 x^{4/b} \left( 1+ \frac{H_2}{x^2} + \mathcal{O}(x^{-4}) \right)  \\
   G(x) &= G_0 x^{-2-2/b} \left( 1+ \frac{G_2}{x^2} + \mathcal{O}(x^{-4}) \right) ,
\end{align}
where the coefficients are
\begin{gather}
   H_0 = \frac{L_0^8 \kappa^4 r_q^4 q_0^4 e^{-4\phi_0}}{b^{12}}, \quad G_0 = \frac{L_0^4}{b^2 r_q^2 \kappa^2} \\
   H_2 = 2(h_2 + 4 f_2 + 2 g_2 - 2 \phi_2), \quad G_2 = h_2 + 2 g_2 \ .
\end{gather}
Now the UV perturbation $\delta S$ is given by
\begin{align}
   \delta S &= \frac{V_6 L_y}{2 G_N^{(N)}} \sqrt{G_0 H_0} \int_{x_*}^\infty \frac{(G_2+H_2)x^{4/b}+(H_2 x^2-(G_2+2 H_2)x_*^2)x_*^{4/b}}{2 x_*(x^{4/b}-x_*^{4/b})^{3/2}} x^{-3+3/b} dx \nonumber \\
   &= \frac{V_6 L_y}{2 G_N^{(N)}} \frac{\sqrt{G_0 H_0}x_*^{-2+1/b}}{2} \int_1^\infty \frac{(G_2+H_2)z^{4/b}+H_2(z^2-2)-G_2}{(z^{4/b}-1)^{3/2}} x^{-3+3/b} dz \nonumber \\
   &= \frac{V_6 L_y}{2 G_{10}} \frac{b \sqrt{G_0 H_0}\sqrt{\pi}}{16} x_*^{-2+1/b} \left( \frac{(2 G_2 + (1+2b)H_2)\Gamma \left(-\frac{1}{4}+\frac{b}{2}\right)}{\Gamma\left(\frac{1}{4}+\frac{b}{2}\right)} - \frac{H_2 \Gamma\left(-\frac{1}{4}\right)^2}{4\sqrt{2}\pi} \right) . \label{deltaS_xs}
\end{align}
We must re-express the above result in terms of the strip width $l$. The strip width near the UV is
\be
   l = 2 \sqrt{H_*} \int_{x_*}^\infty \frac{\sqrt{G(x)}}{\sqrt{H(x)-H_*}} dx = \frac{2 \sqrt{G_0}}{x_*^{1/b}} \frac{b\sqrt{2}\pi^{3/2}}{\Gamma \left( \frac{1}{4} \right)^4} + \delta l \ ,
\ee
where the first term on the last expression is the UV width and $\delta l$ is the first correction. The correction is
\begin{align}
   \delta l &= \sqrt{G_0} x_*^{-2+2/b} \int_{x_*}^\infty \frac{x^{4/b}(H_2 x^2+(G_2-H_2)x_*^2)-G_2 x_*^{2+4/b}}{(x^{4/b}-x_*^{4/b})^{3/2}} x^{-3-1/b} dx \nonumber \\
   &= \sqrt{G_0} x_*^{-2-1/b} \int_1^\infty \frac{z^{4/b}(G_2+H_2(z^2-1))-G_2}{(z^{4/b}-1)^{3/2}} z^{-3-1/b} dz \nonumber \\
   &= \frac{b \sqrt{G_0}\sqrt{\pi}}{8} x_*^{-2-1/b} \left( \frac{(2 G_2+(1+2b)H_2)\Gamma\left(\frac{3}{4}+\frac{b}{2}\right)}{\Gamma\left(\frac{5}{4}+\frac{b}{2}\right)} - \frac{H_2 \Gamma\left(-\frac{1}{4}\right)^2}{4\sqrt{2}\pi}\right).
\end{align}
At this point we know $l(x_*)$. We still need to revert this relation for use in equation (\ref{deltaS_xs}). $x_*$ as a function of $l$ is
\begin{align}
   x_*^{1/b} &= \frac{2\sqrt{2}b\sqrt{G_0}\pi^{3/2}}{\Gamma\left(\frac{1}{4}\right)^2} \frac{1}{l} + \delta x_*^{1/b} \\
   \delta x_*^{1/b} &= \frac{b^{1-2b} G_0^{1/2-b} \Gamma\left(\frac{1}{4}\right)^{4b}}{2^{3+3b}\pi^{-1/2+3b}} l^{-1+2b} \left( \frac{(2 G_2+(1+2b)H_2)\Gamma\left(\frac{3}{4}+\frac{b}{2}\right)}{\Gamma\left(\frac{5}{2}+\frac{b}{2}\right)} - \frac{H_2 \Gamma\left(-\frac{1}{4}\right)^2}{4\sqrt{2}\pi} \right) .
\end{align}
Using these results we conclude that the regulated entanglement entropy on a strip including the first UV correction is
\begin{align}\label{eq:SUV}
   \frac{S(l)}{L_y} = \frac{4\pi^2 F_{UV}(\mathbb{S}^3)}{\Gamma\left(\frac{1}{4}\right)^4} \left(- \frac{1}{l} + \beta(b) l^{-1+2b} + \ldots \right),
\end{align}
where
\begin{align}
   \beta(b) = \frac{(b-1)(-5+3b)\Gamma\left(\frac{1}{4}\right)^{2+4b}\Gamma\left(-\frac{1}{4}+\frac{b}{2}\right)}{2^{7/2+4b} b (5-4b)^b \pi^{1+5b}((2-b)b/\kappa)^{2b}\Gamma\left(\frac{5}{4}+\frac{b}{2}\right)} \ .
\end{align}

One application of this UV expansion is to the determination of the region in the $\{\hat\epsilon, s/l\}$ plane where the theory shows greater super-extensivity than  in a CFT.   As described in Sec.~\ref{sec:extensivity}, we need to compute the derivative of the extensivity parameter $e$ (\ref{eq:ext_definition}) in the UV limit of vanishing strip widths $l$, and in a scaling limit where $s/l$ is held fixed.  Because both $s$ and $l$ are small in this limit, the holographic minimal surfaces computing entanglement entropy are all localized in the deep UV region. The computation of $\partial_l e(l)|_{l=0}$ then involves derivatives of (\ref{eq:ext_definition}), in terms of the expression for entanglement entropy given in  (\ref{eq:SUV}).  This expression is valid for all $\hat\epsilon$, {\emph{i.e.}}, we do not need to take the few flavor limit.

\end{document}